\documentclass[sigconf, nonacm, pdfa]{acmart}

\newcommand\vldbdoi{10.14778/3681954.3682018}

\newcommand\vldbavailabilityurl{https://github.com/tapansriv/arachne}
\newcommand\vldbpagestyle{plain}

\usepackage[utf8]{inputenc}
\usepackage[USenglish]{babel}
\usepackage[a-2b]{pdfx}
\usepackage{colorprofiles}
\usepackage{sparklines}
\usepackage{hyperref}
\hypersetup{pdfstartview=}

\usepackage[export]{adjustbox}
\usepackage{subfig}
\usepackage{xspace}
\usepackage{multirow,graphicx,array}
\usepackage{xcolor}
\usepackage{soul}
\usepackage{marginnote}
\usepackage{enumitem}
\usepackage{amsmath}
\usepackage[noend,linesnumbered]{algorithm2e}
\usepackage{balance}
\newcommand{\eg}{e.g.,\ }

\newcommand{\Lag}{\mathcal{L}}

\newcommand{\tinyskip}{\vspace{3pt}}
\newcommand{\mypar}[1]{\tinyskip\noindent\textbf{#1.}\xspace}

\newenvironment{myitemize}{%
\begin{itemize}[leftmargin=1em, itemsep=.1em, parsep=.1em, topsep=.1em,
    partopsep=.1em]}
{\end{itemize}}

\newenvironment{myenumerate}{%
\begin{enumerate}[leftmargin=1em, itemsep=.1em, parsep=.1em, topsep=.1em,
    partopsep=.1em]}
{\end{enumerate}}

\newenvironment{structure*}{\color{blue}\begin{myenumerate}}{\end{myenumerate}}

\sethlcolor{yellow}

\newcommand{\sys}{\textsf{Arachne}}

\DeclareMathOperator*{\argmax}{arg\,max}

\begin{document}

\title{Saving Money for Analytical Workloads in the Cloud}

\author{Tapan Srivastava}
\affiliation{%
  \institution{The University of Chicago}
}
\email{tapansriv@uchicago.edu}

\author{Raul Castro Fernandez}
\affiliation{%
  \institution{The University of Chicago}
}
\email{raulcf@uchicago.edu}

\renewcommand{\shortauthors}{}


\begin{abstract}
As users migrate their analytical workloads to cloud databases, it is becoming
just as important to reduce monetary costs as it is to optimize query runtime. In
the cloud, a query is billed based on either its compute time or the amount of
data it processes. We observe that analytical queries are either compute- or
IO-bound and each query type executes cheaper in a different pricing model.  We
exploit this opportunity and propose methods to build cheaper execution plans
across pricing models that complete within user-defined runtime constraints.  We
implement these methods and produce execution plans spanning multiple pricing
models that reduce the monetary cost for workloads by as much as 56\%.
We reduce individual query costs by as much as 90\%. The prices chosen by cloud
vendors for cloud services also impact savings
opportunities. To study this effect, we simulate our proposed methods with 
different cloud prices and observe that multi-cloud savings are robust to
changes in cloud vendor prices. These results indicate the massive opportunity
to save money by executing workloads across multiple pricing models.
\end{abstract}

\maketitle

\pagestyle{\vldbpagestyle}

\section{Introduction}
\label{sec:intro}

As analytic workloads migrate to cloud data warehouses, users care just as much
about saving money as they do about optimizing workload runtime.  Even
modest savings on one workload become significant over time if the workload
runs repeatedly.  For example, saving \$140 on an analytics workload that runs
twice a day to update recommendations will save \$100,000 a year, and
organizations may have many such workloads that power different applications
such as filling dashboards to visualize complex data patterns or managing ETL
pipelines~\cite{olap_aws, olap_azure, snowflake_etl_motiv,
goff_blog,sql_etl_blog, google_etl_motiv, dashboard_use_case}. 

While cloud providers offer many tools to tune database performance, there are
no mechanisms to directly save money. Thus, users are increasingly employing
setup-specific solutions to save costs such as working with consulting groups
like McKinsey, the DuckBill group, or CloudZero~\cite{duckbill, cloudzero,
mckinsey} to tune databases, turn off unused resources, and optimally utilize
allocated resources.

\begin{figure}
    \centering
    \includegraphics[width=\linewidth]{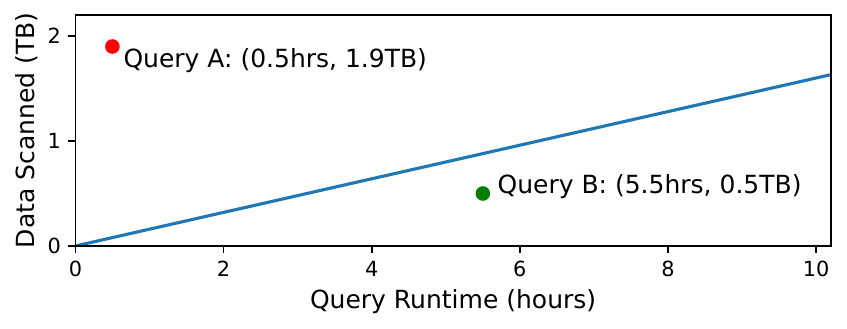}
    \caption{Size scanned (TB) vs. runtime (hours). 
    Boundary for \$6.25/TB (pay-per-byte) vs. \$1/hour
    (pay-per-compute).}
    \label{fig:opp}
    \Description{A blue line with slope of 0.16 separates a red point labelled Query A at coordinates (0.5 hours run, 1.9TB scanned) and a green point labelled Query B at coordinates (5.5 hours run, 0.5 TB scanned).}
\end{figure}


In this paper, we identify opportunities to reduce the monetary cost of running
analytical workloads in the cloud. The key insight is simple. Queries consume IO
\emph{and} CPU; but cloud databases' pricing models charge for IO \emph{or} CPU
time, to keep pricing simple. This opens an opportunity to save money by
cleverly scheduling queries in a cloud database with a favorable pricing model.



The two most prominent pricing models for cloud databases are \emph{pay-per-compute} and
\emph{pay-per-byte}. In a \emph{pay-per-compute} model the user pays for
computation time, \eg in AWS Redshift~\cite{redshift2}, while in
\emph{pay-per-byte} the user pays for the amount of data scanned irrespective of
compute time, \eg in Google BigQuery~\cite{bigquery}. 

CPU-bound queries execute cheaper in pay-per-byte models, vice versa
for IO-bound queries. Figure~\ref{fig:opp} plots query runtime
(hours) on the $x$ axis and the amount of data scanned (terabytes) on
the $y$ axis. We consider one cloud database charging \$6.25/TB (pay-per-byte)
and one charging \$1/hour (pay-per-compute).
Queries on the blue line cost the same in both databases, e.g. one that
runs for 6.25 hours and scans 1TB. 
Query A runs fast but reads 1.9TB (e.g., a simple scan query). It costs
less in the \emph{pay-per-compute} database, so it is above the 
line. Query B scans 0.5TB but runs slower (e.g., window 
operations), so it costs less in the \emph{pay-per-byte}
database. Running each query in the pricing model most beneficial for it is
cheaper than running both queries in a single cloud database.

All workloads have runtime constraints, even if they are loose. For
example, a user with a nightly workload that usually finishes by 2 am may
prefer to delay the completion time up to 8am if they can save 
money~\cite{bq_reliability}. This distills the problem we explore in this paper:
given an analytical workload (a set of queries and data) and a workload runtime
constraint, we propose two algorithms to exploit money saving opportunities
that arise from the observations above:





\begin{myitemize}
\item \textbf{O1: Inter-Query Algorithm.} We propose an algorithm that takes a
    set of queries, data, a workload runtime constraint, and a set of 
    cloud databases and identifies which queries should execute in which 
    databases to save money within the runtime constraint. 

\item \textbf{O2: Intra-Query Algorithm.} We propose an algorithm that, given a
    \emph{single} query, a query runtime constraint, and a set of target cloud databases
    finds subqueries to execute in each cloud database to reduce query cost
    within the runtime constraint. 

\end{myitemize}

Where these complementary techniques are best applied is workload dependent,
e.g., a workload with 3 expensive queries may benefit more from \textbf{O2} than
\textbf{O1}, so we consider each in isolation. However, both \textbf{O1} and
\textbf{O2} require moving data, ensuring cloud database SQL syntax
compatibility, and, more importantly, managing the costs of data movement.  To
exploit \textbf{O1} and \textbf{O2} without modifying user setups, we implement
middleware called \sys{} between users and the cloud that takes a workload and
runtime constraint, executes \textbf{O1--O2}, migrates data, and yields cheaper
execution plans when these exist under the runtime constraint.


We use \sys{} to study the impact of \textbf{O1--O2} on workload costs under
runtime constraints.  We build execution plans across Amazon Redshift
(pay-per-compute)~\cite{redshift2} and Google BigQuery
(pay-per-byte)~\cite{bigquery} to evaluate \sys{}, and we carefully configure
these systems to the best of our ability~\cite{redshift_opt, redshift_best,
bq_opt, bq_best}.

Our results show that there are massive opportunities for saving money.
We achieve up to 57\% savings (we run workloads for \$104 while the original
cost \$243) with an inter-query plan across pricing models that has a nearly 10
hour slowdown. On another workload, an inter-query plan saves 55\% with a 3-hour
speedup. We also achieve up to 90\% savings on a query via an
intra-query plan.

We simulate varying cloud prices and study multi-cloud
savings and runtime-cost tradeoffs as prices change. We see that savings are
robust to price changes and that varying egress fees--what cloud vendors
charge to move data out of a cloud--can aid or fully prevent data
movement. In summary, the contributions of this paper are:

\begin{myitemize}
\item We observe cost saving opportunities for analytical workloads by exploiting different cloud pricing models.
\item We design two algorithms to exploit those opportunities.
\item We implement the algorithms on a system called \sys{}, to realize savings for real workloads.
\item We evaluate \sys{} (and the algorithms it implements) using prominent
    cloud databases and provide a simulation to shed light on
    the impact of market prices on saving opportunities.
\end{myitemize}


Next, Section~\ref{sec:background} presents background and the cost saving
opportunity.  Sections~\ref{sec:inter} and \ref{sec:intra} present the inter-
and intra-query algorithms to exploit \textbf{O1} and \textbf{O2}, and
Section~\ref{sec:system} presents \sys. We evaluate savings opportunities in
Section~\ref{sec:eval} and present related work
in Section~\ref{sec:relatedwork} and conclusions in
Section~\ref{sec:conclusion}.

\section{Background and Opportunities}
\label{sec:background}

We provide background on analytical workloads in the cloud in Section~\ref{sec:executing}, characterize the savings opportunity in Section~\ref{sec:opportunity}, and present the problem statement in Section~\ref{sec:problemstatement}.


\subsection{Executing Analytics Workloads in the Cloud}
\label{sec:executing}

\subsubsection{Cloud Data Warehouses and Pricing Models}
\label{subsec:pricingmodels}

We differentiate \emph{infrastructure-as-a-service} (IaaS) from
\emph{platform-as-a-service} (PaaS) and within PaaS, we separate \emph{pay-per-compute} from \emph{pay-per-byte}.

\mypar{IaaS vs PaaS} In IaaS, OLAP databases (\eg Trino~\cite{presto_paper}
or Apache Hive~\cite{hive_paper}) are manually
deployed and maintained on virtual machines and
billed by compute time. In contrast, PaaS (\eg Google
BigQuery~\cite{bigquery}, AWS Redshift~\cite{redshift2}, Microsoft Azure
Synapse~\cite{synapse}, or Snowflake~\cite{snowflake}) 
deploy and maintain databases for users to directly use. PaaS 
charges more than IaaS for these services.

\mypar{Pay-per-compute vs Pay-per-byte} Two of the most common PaaS pricing models
are \emph{pay-per-compute}, which charges for the amount and duration of
computing resources, and \emph{pay-per-byte}, which charges for the
bytes read by a query regardless of runtime. 


\begin{table}
\caption{Prices for pay-per-compute (PPC), pay-per-byte (PPB),
    storage, and egress. 
    Systems in evaluation are bolded.
}
\centering
\resizebox{0.88\columnwidth}{!}{
\begin{tabular}{|c|c|c|}
\hline
Pricing Model & Database                                & Cost           \\ \hline\hline
\textbf{PPC}   & \textbf{Amazon Redshift--ra3.xlplus}    & \textbf{\$1.086/hr}  \\ \hline 
PPC            & Amazon Redshift--ra3.4xlarge             & \$3.26/hr   \\ \hline 
PPC            & Azure Synapse 100 DWU                   & \$1.20/hr \\ \hline
PPC            & Azure Synapse 500 DWU                   & \$6/hr \\ \hline
PPC            & Snowflake Small (AWS US-East) & \$4/hr \\ \hline  
PPC            & n2-standard-32 VM (GCP)                 & \$1.55/hr \\ \hline
PPC/PPB        & Amazon Redshift Spectrum                & RS + \$5/TB  \\ \hline
\textbf{PPB}         & \textbf{Google BigQuery}         & \textbf{\$6.25/TB} \\ \hline
PPB            & Amazon Athena                           & \$5/TB\\ \hline
PPB            & Azure Synapse Serverless                & \$5/TB\\ \hline
\end{tabular}}
\smallskip
\resizebox{\columnwidth}{!}{
\begin{tabular}{|c||c|c|c|c|}
\hline
Cloud Vendor   & Storage       & Writes          & Reads           & Egress  \\ \hline\hline 
GCP (us-east1) & \$0.023/GB-mo & \$0.05/10k ops  & \$0.004/10k ops & \$120/TB \\ \hline
S3 (us-east)   & \$0.023/GB-mo & \$0.05/10k ops  & \$0.004/10k ops & \$90/TB \\ \hline 
Azure          & \$0.018/GB-mo & \$0.065/10k ops & \$0.005/10k ops & \$87/TB \\ \hline  

\end{tabular}
}
\label{tbl:misc_prices}
\end{table}

\subsubsection{Breakdown of Cloud Costs} 
\label{subsec:breakdown-costs}

Loading data into a cloud database and executing a workload incurs the following costs:

\begin{myitemize}
\item \textbf{Blob Storage cost}: Most cloud databases access data from blob
    storage (\eg AWS S3), and storing data there has a cost (see
    Table~\ref{tbl:misc_prices}). In S3 (us-east) it costs \$23/month to store 1TB of
    data. 
\item \textbf{Read/Write cost}: Blob storage systems charge for API calls, e.g.,
    read/write calls to store and retrieve data from blob storage. For instance,
    it costs \$0.05 to perform 10,000 write operations in S3.
\item \textbf{Loading cost}: While data is loaded into a machine, such a machine
    must be operative and thus consuming compute resources.
\item \textbf{Egress cost}: Transferring data out of a cloud or between regions
incurs per-byte charges. A 1TB transfer from GCP costs \$120.  
\item \textbf{Query Processing cost}: Query execution 
    can be billed per-byte or per-compute. BigQuery charges \$6.25/TB
    scanned.
\end{myitemize}


Running an analytics workload in the cloud involves four steps. In \textbf{Step
1}, data is collected from sources (\eg on-premise repositories, sensors) and
moved to the cloud, incurring data transfer and storage costs. In \textbf{Step
2}, data is loaded into a cloud database incurring read/write and loading
costs. Moving data between clouds 
exacerbates these costs, so our algorithms account for this potential cost increase. In
\textbf{Step 3}, users pay to execute queries against the data in the cloud
database. 
Finally, in \textbf{Step 4} query results are returned to users to use in
downstream tasks, e.g., reporting, filling dashboards~\cite{bq_dashboard,
aws_dashboard}, potentially incurring egress costs. While all four steps incur
costs, costs for \textbf{Steps 1 and 4} depend on the input and output data,
which are mostly fixed for a given workload. We concentrate on \textbf{Steps 2
and 3} which involve cloud databases and dominate the total cost. 
We specifically focus on reducing the cost of \textbf{Step 3} and, to the
extent that data movement is needed, \textbf{Step 2}.



\subsection{Cost Saving Opportunity and Challenges}
\label{sec:opportunity}

We now exploit the insight that migrating CPU- or IO-bound queries or subqueries
to an analytical system with a \emph{beneficial} pricing model presents an cost
saving opportunity.

Given the size scanned by a query ($S$), query runtime ($R$), per-byte cost
($\alpha_S$), and per-compute cost ($\alpha_R$), we observe that $\alpha_S \times S
= \alpha_R \times R \implies S = \frac{\alpha_R}{\alpha_S}\times R$. So a query
that runs for $R$ seconds costs the same in pay-per-compute as one that
reads $\frac{\alpha_R}{\alpha_S}\times R$ bytes in pay-per-byte. This equation
represents the blue line in Figure~\ref{fig:opp} that 
delineates the most \emph{beneficial} pricing model for a query.

\sys{} needs query runtimes to exploit
savings opportunities within runtime constraints.
Unfortunately, there are few accurate approaches to estimating query runtime ($R$)
Instead, Arachne collects query runtime via a
\emph{\textsf{profiling}} stage described in
Section~\ref{subsec:profiler}.

\mypar{Adapting to Cloud Vendor Pricing} This analysis only
    requires query cost and runtime, so \sys{} can support other
pricing models. For tiered pricing models--such as egress where in AWS the
first 10TB/month cost \$90/TB and the next 40TB/month cost \$85/TB--\sys{} can
track usage and adjust pricing constants accordingly.

\sys{} must also track cloud price changes 
    to keep its analysis accurate, as users do
    today. However, pricing changes happen rarely and are announced well in
    advance. Google announced a recent BigQuery price increase 3 months in
advance~\cite{bq_price_change} while Redshift pricing for current
generation hardware has not changed for years.


%
%

\subsection{Problem Statement and Approach}
\label{sec:problemstatement}





We now formally present the goal of this work. Given a workload of queries $Q$
and tables $T$ that execute on a source execution backend $X_s$ under a 
runtime constraint, we consider a set of execution backends (each of which may
offer a different pricing model) and 
find inter- and intra-query plans
(\textbf{O1--O2}) that save money within the runtime constraint.
Users choose which algorithms to
  run on their workload, as \textbf{O1} and \textbf{O2} do not need
  to be deployed together.

\mypar{Approach} To solve the problem statement, we build a proof-of-concept,
\sys{}, which implements the inter- and intra-query algorithms and shows
empirically that it is possible to save money on analytical workloads by
scheduling queries and subqueries across clouds. \sys{} does not require
modifications to existing setups, e.g., where data is stored, and it
handles all needed data movement. \sys{} relies on an offline \emph{profiling}
stage to gather query plans, runtimes, and costs to
save money and meet runtime constraints.

We note that these algorithms can only honor runtime constraints up to the
accuracy of the profiles, \eg if cloud databases dramatically vary a query's
runtime between iterations, the algorithms cannot compute accurate runtime or
cost estimates for plans. We discuss this further with profiling in
Section~\ref{subsec:profiler}.


\section{Inter-Query Execution Plan}
\label{sec:inter}

We now present the inter-query algorithm to exploit \textbf{O1}. We discuss the
setup (Section~\ref{subsec:inter-setup}) and algorithm intuition
(Section~\ref{subsec:greedy}), present the algorithm
(Section~\ref{subsec:inter-alg}), and compare it to an optimal inter-query
algorithm (Section~\ref{subsec:optimal}) to understand its quality.

\subsection{Algorithmic Setup and Goal}
\label{subsec:inter-setup}

We consider an analytics workload with a set of tables $T$, queries $Q$, and
a workload runtime constraint. We assume that initially a user employs a
\emph{source execution backend} $X_s$, paying storage costs for $T$ in $X_s$,
and paying execution costs for $Q$ in $X_s$.

Given a second execution backend, $X_d$, the algorithm chooses a subset of
queries $W \subseteq Q$ to execute in $X_d$ such that the overall workload costs
are reduced without breaking the runtime constraint. To run a query in $X_d$,
all tables that the query scans must migrate from $X_s$ to $X_d$, so the
algorithm must account for migration costs. The algorithm requires the following
inputs:

\begin{myitemize}
\item The set of tables $T$ and queries $Q$ in the workload.
\item $DEADLINE$ is an optional workload runtime constraint.
\item The source $X_s$ and destination $X_d$ execution backends \eg AWS Redshift or Google
    BigQuery.
\item A set of \emph{cloud prices} P = $(p_{blob}, p_{read}, p_{write}, p_{sec},
    p_{byte}$).  $p_{blob}$ per-byte for blob storage, $p_{read}$ read and
    $p_{write}$ write cost, and execution backend costs $p_{sec}$ for
    per-compute pricing models and $p_{byte}$ for per-byte pricing models (example prices
    in Table~\ref{tbl:misc_prices})
\item The egress cost $e$ to move from $X_s$ to $X_d$ \eg \$90/TB out of AWS.
\item A function $s$ which returns the size of a given table. This is 
    measured via cloud storage APIs and is defined for the sake of notation.
\item Functions $C_{X_i}$ and $R_{X_i}$ which take a query $q$ and return the cost and runtime respectively of $q$ in an execution backend $X_i$.
\end{myitemize}

All the above inputs are easy to obtain except for $C_{X_i}$ and $R_{X_i}$,
which are obtained during a profiling stage, explained in
Section~\ref{sec:system}. 
We now formalize the algorithm's goal.

\mypar{Considering the Problem as a Bipartite Graph} We construct a bipartite
graph $G = (T, Q, E)$, with tables $T$ and queries $Q$. We draw an edge $(t\in
T, q\in Q) \in E$ if query $q$ scans base table $t$. 

We next assign weights $\sigma_q$ to each query $q\in Q$ and $\mu_t$ for each
table $t\in T$. $\sigma_q$ represents the \emph{query savings} 
achieved by moving query $q$ to the other execution backend, i.e.,
\begin{equation}
\sigma_q = C_{X_d}(q) - C_{X_s}(q)
\label{eq:sigma}
\end{equation}
$\mu_t$ represents the \emph{migration costs} for a table, which is
the cost of moving $t$ from $X_s$ to $X_d$, loading $t$ into
$X_d$, reading and writing $t$ from blob storage, and temporarily storing $t$ in
blob storage. If each table requires $K$ read/write operations, we can
express $\mu_t$ as:
\begin{equation}
    \mu_t = e \times s(t) + (p_{read} + p_{write}) \times (s(t)/ K) + p_{blob} \times s(t)
\label{eq:mu}
\end{equation}

In Figure~\ref{fig:inter_diagram} we show an example of this model with three
tables $T = \{t_1, t_2, t_3\}$ and three queries $Q = \{q_1, q_2, q_3\}$.
We draw edges to represent query dependencies \eg $q_3$ scans tables $t_2, t_3$. 

The algorithm's goal is to find a subset of tables that maximizes
\emph{query savings}. Concretely, for 
$S\subseteq T$ let $N(S)$ be the set of queries scanning tables in $S$ and let
$N^{-1}(q\in Q)$ be the set of tables $q$ scans. 
Our goal is to find $S_\theta$ in Equation~\ref{eq:opt}.
For example, in Figure~\ref{fig:inter_diagram} we move tables $t_2, t_3$ and
queries $q_2, q_3$ to $X_d$, saving (3+4)-(2+4)=\$1. 

\begin{equation}
S_\theta = \argmax_{S \subset T}\sum\limits_{q \in N(S)}{\sigma_q} -  \sum\limits_{s \in S}{\mu_s}
\label{eq:opt}
\end{equation}


\subsection{Inter-Query Algorithm}
\label{subsec:designing}
Now we provide the intuition for our greedy strategy to exploit \textbf{O1}
(Section~\ref{subsec:greedy}), and present the inter-query algorithm
(Section~\ref{subsec:inter-alg}). Finally, we assess the quality of the greedy
algorithm by comparing it to an optimal (but much slower) min-cut algorithm
(Section~\ref{subsec:optimal}).

\subsubsection{Intuition for the Greedy Strategy}
\label{subsec:greedy}

At each iteration, the greedy algorithm computes the maximum savings achievable
(upper bound) by moving queries depending on $t$ to $X_d$ for each table $t$. It
removes the table with the least upper bound and records the cost and runtime of
the resulting inter-query plan. When no tables remain, it chooses the
cheapest plan with runtime under the runtime bound. 



Concretely, we define $v_t = (\sum_{q \in N(t)}{\sigma_q}) - \mu_t$ as
the sum of \emph{query savings} for all queries that scan $t$ minus the
\emph{migration cost} of $t$. As an upper bound on savings, if $v_t < 0$ it will
\emph{never} be beneficial to move $t$ to the destination backend, so  we
remove nodes for $t$ and all queries scanning $t$ from the bipartite graph.

\begin{figure}
    \centering
    \includegraphics[width=0.85\columnwidth]{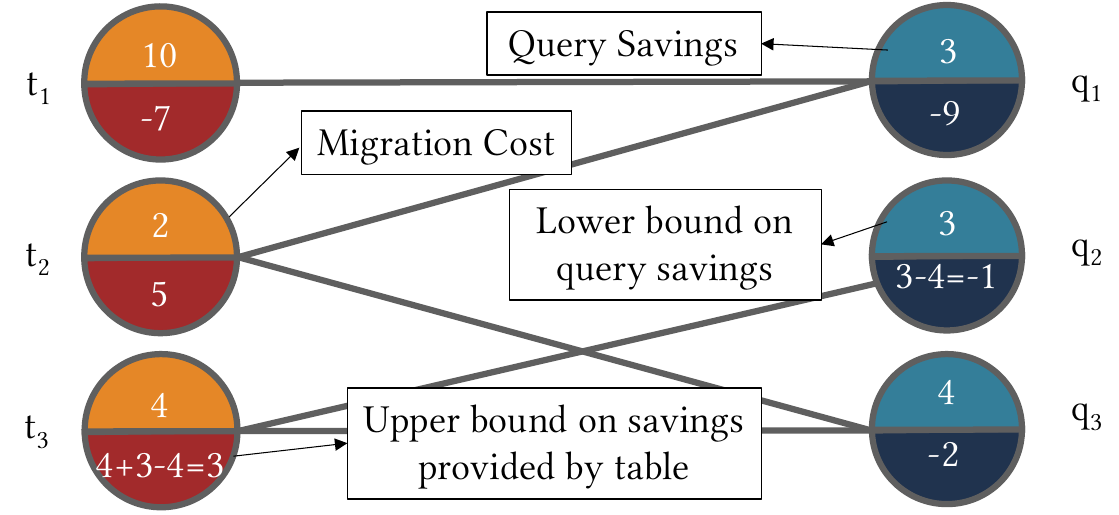}
    \caption{Bipartite model. Node top is query savings $\sigma_q$ or migration
    cost $\mu_t$. Node bottom is upper bound $v_t$ or lower bound $v_q$. We show
how this value is calculated for ${t_3}$ and ${q_2}$.}
    \label{fig:inter_diagram}
    \Description{Bipartite graph. Left side has three circles labelled t1, t2, and t3. The top half is colored orange and the bottom half is red. Each half contains a number. Labels indicate that for the left circles, the top half indicates migration cost while the bottom indicates the upper bound on savings provided by the table. The right side has three circles labelled q1, q2, and q3. The top of these is a lighter blue and the bottom is a darker blue. Labels indicates that the top number is query savings and the bottom is the lower bound on query savings. Lines connect left circles to those on the right. The top of t3 has a 4. t3 is connected to q2 and q3. The top of q2 has a 3, the top of q3 has a 4. The bottom half of t3 contains an equation, 4+3-4=3. This adds the tops of q2 and q3 and subtracts the top of t3. An analogous equation is shown on the bottom of q2 which is only connected to t3. This equation shows 3-4=-1. This adds the top of q2 and subtracts the top of t3. }
\end{figure}

Analogously, we define a lower bound on savings generated from a single query,
$v_q$,  as  \emph{query savings} of $q$ minus the \emph{migration costs}
of the tables that $q$ requires, or $v_q = \sigma_q - \sum_{t \in
N^{-1}(q)}{\mu_t}$.  As a lower bound on possible savings for $q$, if $v_q
> 0$ it is strictly beneficial to move $q$ to $X_d$.  To represent this we add
$q$ and $N^{-1}(q)$ to the final set of queries and tables to move to $X_d$, we
remove the nodes representing $q$ and all $t \in N^{-1}(q)$ from the bipartite
graph, and remove all outbound edges from $N^{-1}(q)$. 
In Figure~\ref{fig:inter_diagram} we compute $v_t$ and $v_q$ and present them in
the lower half of each node, \eg $v_{t_2}$ is the savings of $q_1$ and $q_3$
minus the cost of $t_2$, or $3+4-2 = 5$. 

\begin{figure}
    \centering
    \includegraphics[width=0.9\columnwidth]{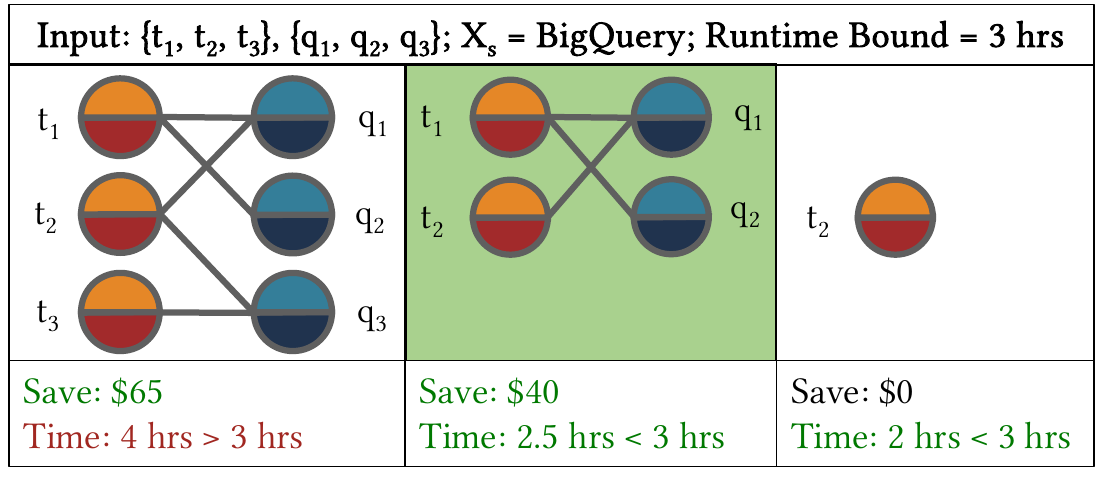}
    \caption{The plans considered with a 3 hour runtime
    bound. Plan B saves the most money within 3 hours so it is chosen.}
    \label{fig:inter_sla}
    \Description{A diagram showing the three plans considered for a problem with three tables as input labelled t1, t2, and t3 and three queries as input q1, q2, and q3. The starting cloud is BigQuery, and the runtime bound is 3 hours. The first plan considers migrating all tables and queries. This saves 65 dollars but runs in 4 hours, which is longer than the runtime bound. The second plan removes t3 and q3. This saves 40 dollars and runs in 2.5 hours, which is under the runtime bound. The final plan only contains t2. This saves 0 dollars and runs in 2 hours, the fastest of the three. The second plan is shaded green, as it saves the most money of the plans that run under 3 hours.}
\end{figure}

\mypar{Algorithm Example} In Figure~\ref{fig:inter_sla} the algorithm considers three
plans for the given workload.
The first plan (left) migrates all tables and queries, saving \$65 but
violating the runtime constraint, so runtime is colored red.
Removing $t_3$ yields the second plan (middle), saving \$40 and running
in 2.5 hours, so both savings and runtime are colored green. Finally, removing
$t_1$ yields the baseline (right) which saves no money. The
second plan is chosen as it saves the most money under the runtime constraint,
so it is colored green. 


\begin{algorithm}
\small
\footnotesize
\LinesNumbered
\SetKwInput{Input}{input}
\SetKwInput{Output}{output}
\SetKwData{KwTo}{to}
\SetKwProg{Fn}{Function}{:}{end}
\SetKwFunction{ProcName}{ReducePlan}
\SetKwFunction{FnName}{InterQuery}
\BlankLine
\Fn{\FnName{$\{X_s, X_d\}$, prices P,  $e$,  $T$, $Q$, $s$, $C$, $R$, $DEADLINE$}} {
    $T', Q', T_f, Q_f$ = ReducePlan(X, P, e, T, Q, s, C) \;
    Remove all outbound edges from $T_f$ \;

    \While{ $T' \neq \emptyset$}{
        For $t \in T'$, $v_t = (\sum_{q \in N(t)}{\sigma_q}) - \mu_t$ \;
        $t' \in T'$ be the table with minimium $v_{t'}$  \;
        $T' = T' - \{t'\}$; $Q' = \{q | N^{-1}(q) \subset T'\}$ \;
        $T', Q', T_f', Q_f' = ReducePlan(X, e, T', Q', s, f)$ \;
        $T' = T' \cup T_f'$; $Q' = Q' \cup Q_f'$ \;
        planCosts[$T', Q'$] = $\langle$cost($T', Q'$), runtime($T', Q'$)$\rangle$ \;
    }

    Return the min cost plan in planCosts within $DEADLINE$ \;
}
\Fn{\ProcName{$X$, P, $e$, $T'$, $Q'$, $s$, $C$}} {
    $Q' = \{q | \sigma_q > 0 \}$ ; $T_f, Q_f = \{\}, \{\}$ \;

    For $t\in T'$ $v_t = (\sum_{q \in N(t)}{\sigma_q}) - \mu_t$ \;
    For $q\in Q'$ $v_q = \sigma_q - \sum_{t \in N^{-1}(q)}{\mu_t} $ \;

    \While{ $T' \neq \emptyset \land \exists v_t < 0 \land \exists v_q > 0$}{
        $U = \{t | v_t < 0 \}$; $V =  \{q | v_q > 0\}$ \;
        $T' = T' - U$ ; $Q' = Q' - \{q | q \in N(t) \forall t \in U\}$ \;
        $T' = T' - \{t | t \in N^{-1}(q) \forall q \in V\}$ ; $Q' = Q' - V$ \;
        $T_f = T_f \cup \{t | t \in N^{-1}(q) \forall q \in V\}$ ; $Q_f = Q_f \cup V$ \;
        For $t\in T'$ $v_t = (\sum_{q \in N(t)}{\sigma_q}) - \mu_t$ \;
        For $q\in Q'$ $v_q = \sigma_q - \sum_{t \in N^{-1}(q)}{\mu_t} $ \;
    }
    return $T', Q', T_f, Q_f$\;
}
\caption{Inter-query greedy algorithm}
\label{alg:inter}
\end{algorithm}
 
\subsubsection{The Inter-Query Algorithm in Depth}
\label{subsec:inter-alg}

The greedy algorithm, shown in Algorithm~\ref{alg:inter}, first invokes
\textsc{ReducePlan} (line 2), which computes $v_t$ and $v_q$ (lines 14--15), 
removes tables with $v_t < 0$ from consideration, and migrates
queries with $v_q > 0$ to $X_d$ (lines 17--22). This loop repeats 
until there are no more candidates or there are no
tables left (line 16). \textsc{ReducePlan} returns the tables and queries to
migrate and the remaining tables and queries to consider (line 23). 

The algorithm then removes all outbound edges from the tables migrating to $X_d$
(line 3) and proceeds to greedily remove tables from consideration. While there
are still tables (line 4), the algorithm computes $v_t$ (line 5),
assigns $t$ with minimal $v_t$ to remain in $X_s$,
and removes nodes in $N(t)$ (line 6--7). The algorithm calls
\textsc{ReducePlan} again to prune away tables with $v_t < 0$ and identify
queries with positive lower bounds (line 8). The algorithm records the cost and
runtime of the current plan using query costs and runtimes $C$ and
$R$, cloud prices $P$, and table sizes $s$ (line 10).

Finally, the algorithm chooses the cheapest plan in $planCosts$ with runtime
less than $DEADLINE$ (line 11). The baseline plan-- migrating no
tables or queries--will be cached in $planCosts$ and will be chosen if no
cheaper plans exist under the runtime constraint.

\subsubsection{Optimal Algorithm and Complexity Analysis}
\label{subsec:optimal}

To evaluate the greedy algorithm we: i) present an optimal min-cut
based inter-query algorithm; ii) compare its runtime complexity to the greedy
approach; iii) and show the greedy algorithm's accuracy in practice.

\mypar{Optimal Solution}
As in~\cite{rpsp}, we build a capacity
function $c$ and a bipartite graph $G$ with tables on the left 
and queries on the right. Edges
with infinite capacity connect tables and queries by query dependencies.
We add a source node $a$ and draw edges from $a$ to every table $t_i$; $c((a,
t_i)) = \mu_{t_i}$. We add a sink node $b$ and draw edges from every query
$q_j$ to $b$; $c((q_j, b)) = \sigma_{q_j}$. 
The algorithm finds the min-cut $(A,B)$ and migrates the
queries and tables in $B$ to $X_d$.
%
%


\mypar{Complexity Analysis} 
Using a min-cut algorithm~\cite{edmonds_karp, Dinitz2006}, the optimal algorithm
has complexity $O(|V|^2|E|)$. 
For the greedy
approach, (note: $|V| = |T| + |Q|$), computing $v_t$ or
$v_q$ is $O(|V|)$ with at worst $|T|$ iterations, yielding
a worst-case complexity of $O(|T||V|)$. The optimal algorithm
is both an order of magnitude less efficient and depends on the
number of relationships between queries and tables. The greedy algorithm
is independent of query complexity.

\mypar{Practical Significance}
This difference in complexity between the optimal and greedy
    algorithms has practical significance when the number of queries and tables
    is large. For example, when using the TPC-DS workload as input (24 tables
    and 100 queries), the difference is insignificant, with both algorithms
    running in less than 0.3 seconds. With 1000 queries and 100 tables, the
    optimal runtime jumps to 3.4 seconds, while the greedy remains under 0.3
    seconds. With 2500 queries and 400 tables, the optimal algorithm takes 2.1
    minutes while the greedy takes 1.2 seconds.  


\mypar{Greedy Algorithm Accuracy} To evaluate accuracy, we use 72 workloads
at 1TB and 2TB, both with and without IaaS, and using both internal
and external BigQuery storage, producing 576 workloads. We run the
greedy and optimal algorithms on each workload. Our greedy strategy finds the
optimal solution for all workloads.



\section{Intra-Query Execution Plan}
\label{sec:intra}

We now present the intra-query algorithm to exploit \textbf{O2}. We first present the setup
(Section~\ref{subsec:intra-setup}) and how to
identify where to make cuts (Section~\ref{subsec:intra-inputs}) before
presenting the algorithm (Section~\ref{subsec:intra-algo}).  

\subsection{Algorithmic Setup and Goal}
\label{subsec:intra-setup}

The setup is similar to that in Section~\ref{subsec:inter-setup},
but this algorithm takes a single query $q$ and a runtime
constraint for $q$ and
finds a subquery that saves money when migrated from $X_s$ to $X_d$ after
accounting for migration costs while honoring the runtime constraint.

\mypar{Formalization and Goal} A query plan is a directed acyclic graph where
leaves represent base tables and edges represent data flow from upstream tables
to downstream operators.  Removing all outbound edges from a node partitions the
graph into two disjoint subgraphs; we call this process making a \emph{cut} in
the query plan at a node. The goal of the intra-query algorithm is to find a
\emph{profitable} cut
of $q$ so that one subquery executes in $X_s$ and the other in $X_d$ so that
the total query cost, including migration cost, is lower than running the entire
query in $X_s$ while adhering to the runtime constraint. 

\subsection{Identifying Profitable Cuts}
\label{subsec:intra-inputs}
We now present the insight we use to identify profitable cuts.  Let $T = (V, E)$
be a query plan. Let $p_{byte}$, $p_{sec}$, and $e$ be per-byte, per-compute,
and egress prices. Let migration cost $\mu_t$ be as in Equation~\ref{eq:mu} and 
let $\mu_v$ for $v\in V$ be the migration cost of the data output from $v$.
Let $s$ be the table size and $rs(v)$ be the row size for $v$. Finally,
when a cut is made at a node $v$, the subgraph \textbf{u}pstream of $v$ is
$S_{\textbf{u}}(v)$--including $v$ and all base tables that flow into $v$--and
the subgraph \textbf{d}ownstream of $v$ is $S_{\textbf{d}}(v)$. We now show some
key definitions.

\begin{myitemize}
\item $f_w(v)$ returns the output cardinality of $v\in V$
\item $f_r(v)$ returns the runtime of $S_u(v)$
\item $DEADLINE$ is an optional runtime constraint for $q$. 
\item $\Lag(v)$: the base tables in the downstream subquery $S_d(v)$.
\item $C_s(v) = \alpha_s \sum_{u\in \Lag(v)}{f_w(u)rs(u)}$: the cost of $S_d(v)$ per-byte.
\item $C_r(v) = p_{sec} f_r(v)$: the cost of $S_u(v)$ per-compute. 
\item $C_m(v) = \mu_v + \sum_{t \in \Lag(v)}{\mu_t}$: the cost to migrate all necessary data, including the output of $v$, to $X_d$.  
\end{myitemize}

These values require query costs $C_{X_i}$ and runtime $R_{X_i}$ 
in all execution backends. The algorithm's goal is to find $v\in
V$ such that: 
\begin{equation}
C_r(v) + C_m(v) + C_s(v) < C_{X_s}(q) 
\label{eq:intra}
\end{equation}

This finds the cheapest plan, represented
on the left, that costs less than simply executing the query in $X_s$.

\mypar{Insight} Naively, we could find the optimal execution plan 
by making a cut at every operator and executing each resulting
plan. This approach requires we pay query processing and migration costs 
for each possible plan, and the number of possible plans
grows with the size of the query. Our goal is to find cheaper execution
plans while minimizing the incurred overhead costs.

To achieve this goal, we assign a value to each operator in $q$ corresponding to
the \emph{savings opportunity} of making a cut at that operator. Using the inputs to
the algorithm, we reorder Equation~\ref{eq:intra} and
compute the maximum savings achievable by an intra-query plan.
We consider those operators with positive savings opportunity and use
this value to guide what candidate operators we evaluate.

\mypar{Calculating Savings Opportunity} 
The \emph{savings opportunity} $o_v$ is the right hand side of $ p_{sec}f_r(v) <
C_{X_s}(q) - (C_m(v) + C_s(v))$, derived from Equation~\ref{eq:intra}, which we
compute using $C_{X_i}$ and $f_w$.
We draw two conclusions. First, if $o_v < 0$, the plan produced from a cut at
$v$ will cost more than the baseline. Second, the only way to determine if a
plan will cost less is to pay to compute $f_r$, so the algorithm aims to reduce
the number of times it computes $f_r$.

\begin{algorithm}
\LinesNumbered
\small
\footnotesize
\SetKwInput{Input}{input}
\SetKwInput{Output}{output}
\SetKwData{KwTo}{to}
\SetKwProg{Fn}{Function}{:}{end}
\SetKwFunction{FnName}{IntraQuery}
\BlankLine
\Fn{\FnName{$T = (V,E)$, $X$, P, $e$, $f_w$, $C$, $R$, $DEADLINE$}} {
    \For{$u \in V$}{
        $o_u = C_{X_s}(q) - (C_m(u) + C_s(u))$ \;
    }
    $candidates = \{ v \in V | o_v > 0\}$ \;
    \While{$candidates \neq \emptyset$ or number of iters $< K$} {
        Pick $u$ from $candidates$ such that $o_v$ is largest \;
        Compute $f_r(u)$ \;
        Compute $a_u = o_u - p_{sec}f_r(u)$ \;
        \For{$v \neq u \in candidates$}{
            If $o_v < a_u$ remove $v$ from $candidates$ \;
        }
        \For{$v \in candidates$;  $v$ downstream of $u$}{
            $o_v = o_v - p_{sec}f_r(u)$ \;
            If $o_v < 0$ remove $v$ from $candidates$ \;
        }
    }
    Return $u$ with maximum $a_u > 0$ under $DEADLINE$ or baseline\;

}
\caption{Intra-query algorithm}
\label{alg:intra}
\end{algorithm}

\mypar{Updating Opportunities} Measuring $f_r$ allows us to update the
opportunities for other nodes.  For example, if a node $u_1$ sits downstream of
another node $u_2$, $f_r(u_1) \geq f_r(u_2)$, so if we compute $f_r(u_2)$,
$o_{u_1}$ must decrease by $p_{sec}f_r(u_2)$ since it must pay at
least that much in runtime cost. If we measure the savings for a plan $a_u$, we can
remove all candidates with $o_{u_1} < a_{u_2}$ because a cut at $u_1$
cannot produce a cheaper execution plan. We can then remove multiple
candidates per iteration, reducing the number of invocations to $f_r$. 

\subsection{Intra-Query Algorithm}
\label{subsec:intra-algo}
The intra-query algorithm, in Algorithm~\ref{alg:intra}, computes the
opportunity $o_u$ for every node (lines 2-3) and picks
candidates with $o_u > 0$ (line 4). It iterates over the
candidates from largest to smallest $o_u$ (line 6). Computing $f_r$
for all candidates may be expensive, so iterating from largest to
smallest opportunity ensures that the potential savings lost by not iterating
over all candidates are minimized.

For each candidate $u$, the algorithm computes $f_r(u)$, the real savings,
$a_u$, (lines 7--8) and updates the opportunity for other candidates (lines
10--16). 
It repeats this until all candidates have been checked or after $K$ iterations
(line 5). The algorithm chooses the cut that
yields the most savings within the runtime constraint or the baseline if no such cut
exists (line 14). 


\mypar{Complexity and Discussion of Optimality} The algorithm removes
at least one candidate per iteration, so the worst-case complexity is $O(|V|)$.
The algorithm by default considers all cuts in a
query plan that could yield savings and chooses the optimal cut. 
The algorithm only parses a single query plan, 
but it will never choose an
execution plan more expensive than the baseline.

\begin{figure}
    \centering
    \includegraphics[width=\columnwidth]{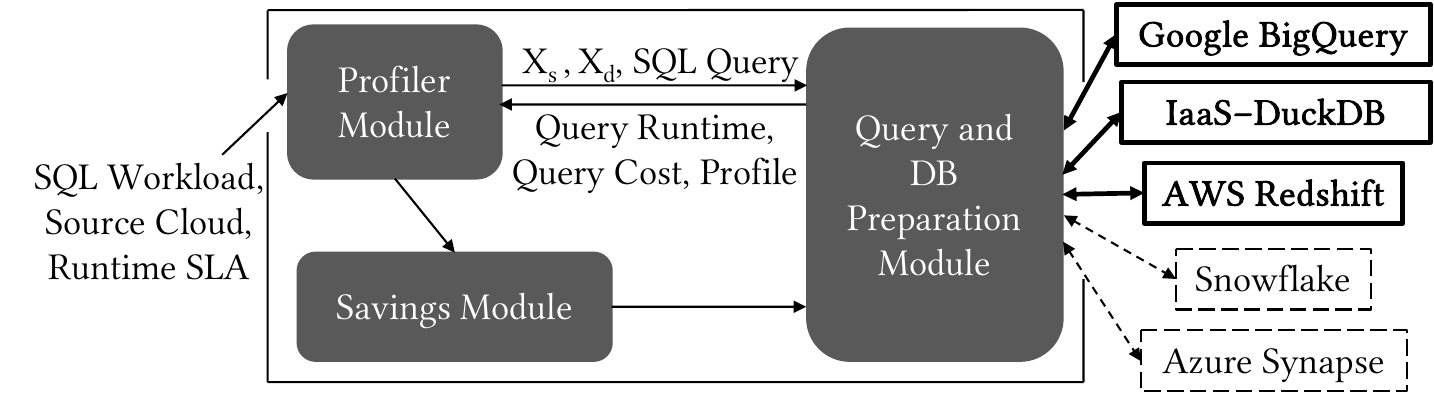}
    \caption{\sys{} architecture and execution backends.}
    \label{fig:system}
    \Description{A box is drawn containing three smaller boxes, labelled Profiler Module, Savings Module, and Query and DB Preparation Module. Lines are drawn from Profile Module to Savings module, from Savings Module to Query and DB Preparation module, and back and forth between Profiler Module and Query and DB Preperation module. The line from the profiler module to the Query and DB Preperation module has a source cloud,a destination cloud and a SQL query. The reverse line has Query Runtime, Query Cost, and a Profile. The box encompassing these three modules has a line incoming to it carrying a SQL workload, a source cloud, and a runtime SLA. The query and DB prepraration module connect to 5 cloud databases. Three are bolded as they are included in the evaluation.}
\end{figure}

\section{\sys{} Overview}
\label{sec:system}

We implement \sys{}, including
the inter- and intra-query algorithms, in 8k lines of Python and Java.
We present \sys{'s}
architecture in Section~\ref{subsec:overview}, the \emph{\textsf{profiler}}
in Section~\ref{subsec:profiler}, and other cost-relevant
implementation decisions in Section~\ref{subsec:implementation}.


\subsection{Overview}
\label{subsec:overview}

Figure~\ref{fig:system} shows \sys{'s} architecture.  Users first execute
\textsc{initialize}, which submits an unmodified SQL
workload, stored as individual SQL files, the source
execution backend where data starts, and the optional runtime
constraint. \sys{} implements the inter- and intra-query algorithms
in the \textbf{\textsf{savings} module}.  The \textbf{\textsf{profiler} module}
gathers the query costs and query runtime inputs ($C_{X_i}$ and $R_{X_i}$ from Sections~\ref{sec:inter}--\ref{sec:intra}) 
so the \textsf{savings} module can save money and meet the runtime
constraint. The \textbf{\textsf{preparation} module} prepares queries for
execution in target backends. It 1) ensures that SQL
queries are compatible with the syntax requirements of execution backends; 2)
submits queries for execution; 3) orchestrates materializing data into
portable, open-format Parquet files; and 4) migrates those Parquet files between
execution backends when needed. 


\mypar{Execution Backends} \sys{} supports two PaaS
backends--AWS Redshift and Google BigQuery--so we can study per-compute and
per-byte pricing models. It can also deploy DuckDB on IaaS. 
These backends are bolded in
Figure~\ref{fig:system}. \sys{} can be easily extended to support new
backends, such as those in dashed lines in Figure~\ref{fig:system}.

\mypar{Minimizing Infrastructure Changes}
\sys{} is designed to be compatible with existing ETL
pipelines (and downstream BI tools and dashboards), so it moves data
between source backends \emph{at runtime} whenever this saves money
and does so transparently to the end user. Consequently, 
\sys{} moves data every time a workload executes (incurring
costs) but does not need to handle data inconsistencies that may
arise if, for example, the data were replicated in several backend systems.
Exploring tradeoffs of the spectrum of solutions (keep ETLs intact
and pay repeated migration costs or change ETLs, duplicate data, and
incur double storage costs) is beyond the scope of this paper.
    Despite minimizing infrastructure changes, \sys{}
    still requires configuration changes, e.g., to access cloud accounts. We
    believe this is not an important roadblock to deployment, as such
    credentials are frequently shared with other tools. These changes add
    one-time costs; in exchange, \sys{} achieves large recurring savings (see
Section~\ref{sec:eval}).


\mypar{Compliance}
At a high level, \sys{} is best seen as an
alternative interface to the source backend. When data (or queries) cannot 
move (or execute) between vendors, e.g., for compliance requirements
preventing data migration to a geographical area, such data can be
excluded from \sys{} and run using the usual interface.


\mypar{Implementation} \sys{}
uses Apache Calcite \cite{calcite} to convert between SQL and query plans and
perform query optimization. \sys{} uses Apache Arrow~\cite{arrow} to sample
tables; the Redshift-Data API to use Redshift clusters and S3; and the
Google Cloud client for Storage and BigQuery. It migrates all data at runtime.



\subsection{\textsf{Profiler} Module}
\label{subsec:profiler}

The \textsf{profiler} module provides the inter- and
intra-query algorithms with accurate runtime, cost, and cardinality information
for each query. We consider two approaches: prediction and profiling.

\mypar{Why We Do Not Use Prediction in \sys{}} Existing
    approaches to estimate operator cardinality $f_w$, query cost $C_{X_i}$, and
    query runtime $R_{X_i}$ are noisy. Cardinality estimates from query
    optimizers remain
    inaccurate~\cite{lohman_solved_2014,yang_deep_2019,perron_how_2019}.
    The ``cost'' produced by query optimizers for a given query plan has only a
    relative meaning to the cost associated to other plans rather than to 
    absolute monetary or runtime cost, making query optimizers' cost a poor
proxy for estimating runtime~\cite{leis_how_2015,wu_cost_model}. Last, there are
few approaches for estimating query runtime given a query plan, and existing
ones often underperform, as we show in Section~\ref{subsec:micro}.
However, we anticipate that continued advances in this area could replace the
profiling approach we now explain.

\mypar{The Profiling Approach} 
The \textsf{profiler} executes each query in
each execution backend, obtaining its runtime and cost. It also gathers the
output cardinality for each query operator via query profiling in
DuckDB, and provides the inputs to the inter- and intra-query algorithms 
($C_{X_i}$, $R_{X_i}$, and $f_w$
as defined in Sections~\ref{sec:inter}--\ref{sec:intra}). 
Profiling is more accurate and more costly than prediction. Profiling cost is
amortized if the profiled queries execute several times without major runtime
changes. This is the case for most periodic workloads~\cite{tan_choosing_2019,
aws_batch_bi, bq_reliability}, which happen to be the most relevant for saving
money. Furthermore, stale profiles can still expose savings opportunities since
small errors in costs do not greatly alter what queries migrate in an
inter-query plan, as we illustrate in Section~\ref{subsec:micro}.


\mypar{Profiling Over Samples}
It is possible to reduce profiling costs by measuring runtime, cost, and operator cardinality on a workload sample and then extrapolating to the original workload size without introducing significant error. It is
difficult to extrapolate query runtime from samples, which is related to the
difficulty of join sampling. If the output of a join over data samples is not a
sample of the true join result, we cannot accurately extrapolate runtime for the
full join~\cite{surajit_sampling}. However, profiling costs are largely in
\emph{pay-per-byte} pricing models and depend only on the data size. 
We show empirically in Section~\ref{subsec:micro} that the
profiling cost is quickly compensated by workload savings, often in less than 5
executions of the cheaper execution plan when profiling over the entire dataset
and in 1--2 iterations when using a sample. For periodic workloads, we can
collect profiles during iterations of the workload to reduce the net profiling
costs. We see then that profiling costs are quickly
compensated by workload savings.

\mypar{Assigning Operator Cardinalities} 
 To provide operator cardinality $f_w$
 (Section~\ref{subsec:intra-inputs}), \sys{} profiles queries in DuckDB as
 cardinality is independent of execution backend. However, DuckDB's physical
 plan may not match \sys{'s} internal query plan, so \sys{} cannot directly
 match operator profiles from DuckDB onto its query plan.
We observed that DuckDB does not re-order operators between SQL subqueries, 
so \sys{} writes its query plan into SQL, with all operator trees as nested
subqueries, so DuckDB's physical plan exactly matches \sys{'s} and \sys{} can
assign cardinalities to operators in its internal query plan. 

%

\subsection{Cost-Relevant Implementation Details}
\label{subsec:implementation}

\mypar{Data Transfer Between Clouds} Cloud vendors provide easy-to-use tools
(\eg AWS DataSync) to transfer data into their clouds. Beyond the
egress cost of doing so, users have to pay for these tools too.
\sys{} implements a simple cloud transfer tool using blob storage
APIs. Our tool on a GCP n2-standard-32 VM transferred a 615GB dataset for
\$0.58; that transfer in AWS DataSync costs \$7.69.

\mypar{SQL Compatibility} \sys{} builds and applies text rules to ensure
\sys{}-produced SQL is compatible with different backend dialect rules,
\eg BigQuery requires column names to contain only alphanumerics or underscores.

\mypar{Calcite Query Operators} 
\sys{} implements its own physical node subclass and uses Calcite libraries to
perform heuristic optimizations like predicate pushdown, make cuts in query
plans, and assign cardinalities collected during profiling to query operators.


\section{Evaluation}
\label{sec:eval}

In this section, we answer the following research questions:

\begin{myitemize}
    \item \textbf{RQ1:} Does the inter-query algorithm save money?  (\textbf{O1})
    \item \textbf{RQ2:} Does the intra-query algorithm save money? (\textbf{O2})
    \item \textbf{RQ3:} How does pricing (chosen by cloud vendors) affect inter-query savings and the runtime-cost tradeoff?
    \item \textbf{RQ4:} 
            \textsf{Profiler} Microbenchmarks: How does sampling impact
            profiling costs and accuracy and how quickly are they compensated?
            Does using stale profiles diminish savings?  How are savings
            impacted by noisy runtime estimates versus profiles?
    
\end{myitemize}

Because \sys{} can deploy DuckDB on IaaS, we also evaluate how utilizing
cheaper IaaS impacts inter-query savings. 

\subsection{Workloads}
\label{subsec:workload}

\mypar{Resource Balance Workloads} The specific \emph{balance} of CPU- and
IO-queries in a workload will impact savings opportunities. We use the well-known
TPC-DS~\cite{tpcds} benchmark to create three workloads each with a different
balance of CPU- and IO-bound queries to explore the design space (in the
original TPC-DS benchmark nearly all of the 99 queries are IO-bound). We adapt
queries from LDBC, a well-known business intelligence benchmark~\cite{ldbc}, 
to work on data from TPC-DS: we create queries to find customers
related to each other by purchase history and queries to find
connected components of customers for recommendation algorithms. We combine the
authored CPU- and some existing IO-bound queries to create three workloads over
17 tables with different characteristics:




\begin{myitemize}
    \item \textsc{W-CPU}: 46 queries, about 40\% of which are CPU-bound.
    \item \textsc{W-Mixed}: 49 queries, about 30\% of which are
    CPU-bound.
    \item \textsc{W-IO}: 46 queries, about 20\% of which are
    CPU-bound.
\end{myitemize}

While there are more IO-bound queries in each workload, the CPU-bound queries
consume a large amount of CPU, so overall resource
consumption for each workload reflects the workload's name. We make all workloads
publicly available~\footnote{https://github.com/tapansriv/resource-balance-workloads}
for reproducibility and because they may be of independent interest to others.


\mypar{Read-Heavy Workloads} We also explore skewed workloads. While nearly all TPC-DS queries
are IO-bound with runtime dominated by table \emph{reads}, these queries 
differ in runtime and complexity. To explore savings opportunities on a range
of IO-bound workloads we create 24 workloads called the Read-Heavy workloads
from TPC-DS, which contains 24 tables and 99 queries, by removing one table
from the TPC-DS dataset. This creates a 23-table dataset with a subset of the
original 99 queries, on average about 80 queries. Each workload is named by
the alphabetical order of the table that was removed to generate it, \eg the
workload created by removing the first table alphabetically,
\textit{call\_center}, will be named \textbf{Read-Heavy 0}.

\mypar{LDBC} Additionally, we use queries written on the LDBC Social Network
Benchmark-Business Intelligence (SNB-BI) dataset~\cite{ldbc}. 

For \textbf{RQ1} and \textbf{RQ3} we use the
Resource Balance and Read-Heavy Workloads. For \textbf{RQ2}, we use
TPC-DS queries and author queries on TPC-DS and LDBC. We explain these in
detail in that section.

\subsection{Experimental Setup}
\label{subsec:setup}

\mypar{PaaS Execution Backends} 
We use Google BigQuery (\emph{pay-per-byte}) and AWS Redshift
(\emph{pay-per-compute}) as popular representatives of each pricing model. In
Redshift cluster size impacts cost and performance, so we explore the
\textsc{G$\rightarrow$A1}, \textsc{G$\rightarrow$A4}, \textsc{G$\rightarrow$A8}
setups where data starts in BigQuery (\textsc{G}) and we consider migrating to a
1-, 4-, or 8-node ra3.xlplus Redshift cluster respectively ($\rightarrow$A1,
$\rightarrow$A4, or $\rightarrow$A8;  the arrow indicates the migration direction). 
We explore the \textsc{A4$\rightarrow$G} setup where
data starts in a 4-node ra3.xlplus Redshift cluster and could migrate
to BigQuery.  We optimize our Redshift and BigQuery setup per docs and best
practices~\cite{redshift_opt, redshift_best, bq_opt, bq_best}.




\mypar{Data Format and Storage} We store intermediate data in Parquet
files~\cite{parquet} with Snappy compression~\cite{snappy}. All cloud databases
we use are compatible with open data formats~\cite{bricks_lake, azure_lake}. We
create external tables in BigQuery pointing to the Parquet files in blob
storage. We also consider data loaded into BigQuery. Redshift loads Parquet
files from S3. The compression of data saves migration costs, and all in-flight
compression occurs during materialization from pay-per-compute databases and is
billed as runtime cost.

\mypar{Where Data Is Initially Stored} 
For the Resource Balance Workloads, we consider G$\rightarrow$A4 and
A4$\rightarrow$G. With current prices, Redshift is significantly cheaper than
BigQuery for IO-skewed workloads and queries. As such, we consider only
G$\rightarrow$A1, G$\rightarrow$A4, G$\rightarrow$A8 for the Read-Heavy
workloads to evaluate \textbf{O1} and only consider DuckDB and BigQuery on data
stored in GCP to evaluate \textbf{O2}, as there are few savings opportunities if data starts in Redshift.


\mypar{How Workloads Are Executed} Batch, analytic workloads like those in this
evaluation are often executed serially but can also be submitted all at once to
a system~\cite{ibm_serial_batch,aws_batch}. For pay-per-byte systems like
BigQuery, this has no impact on workload cost, only runtime. For pay-per-compute
systems like Redshift, this also impacts costs. Redshift's REST API, the
Redshift Data API~\cite{redshift_data_api} offers \textsc{BatchExecuteStatement}
which executes a list of SQL statements one at a time in a single transaction. 



\mypar{Metrics} We measure monetary cost in US dollars and runtime in time
units. We account for all applicable cloud costs (see
Section~\ref{subsec:breakdown-costs}) and use cloud prices as of February'24 in
Table~\ref{tbl:misc_prices}. We validate our results by checking the breakdown
of charges in our account from cloud vendors. We create VMs in
the same cloud region as blob storage buckets to avoid regional transfer
costs and utilize data compression to reduce migration costs.

\mypar{Runtime Constraint} We assume that there are no
runtime constraints and concentrate on exploring cost savings without 
imposing arbitrary runtime constraints that do not add any additional insight.

\subsection{RQ1. Inter-Query Processing}
\label{subsec:inter-query}

We now explore \textbf{O1}. First, we evaluate the cost opportunities of the
inter-query algorithm across the three Resource Balance Workloads
(Section~\ref{subsubsec:mcw}) 
and
the IO-skewed Read-Heavy workloads (Section~\ref{subsubsec:ds-derived}). Then,
we leverage that \sys{} can deploy DuckDB on IaaS to evaluate how that impacts
savings (Section~\ref{subsubsec:iaas}).

\subsubsection{Resource Balance Workloads}
\label{subsubsec:mcw}

We run the inter-query algorithm on \textsc{W-CPU}, \textsc{W-IO}, and
\textsc{W-Mixed} in G$\rightarrow$A4 and A4$\rightarrow$G and evaluate
multi-cloud savings. First, we provide an overview of the results before presenting
an in-depth breakdown of costs.

\begin{figure}%
    \centering
    \subfloat[1TB A4$\rightarrow$G]%
    {{\includegraphics[width=0.5\columnwidth]{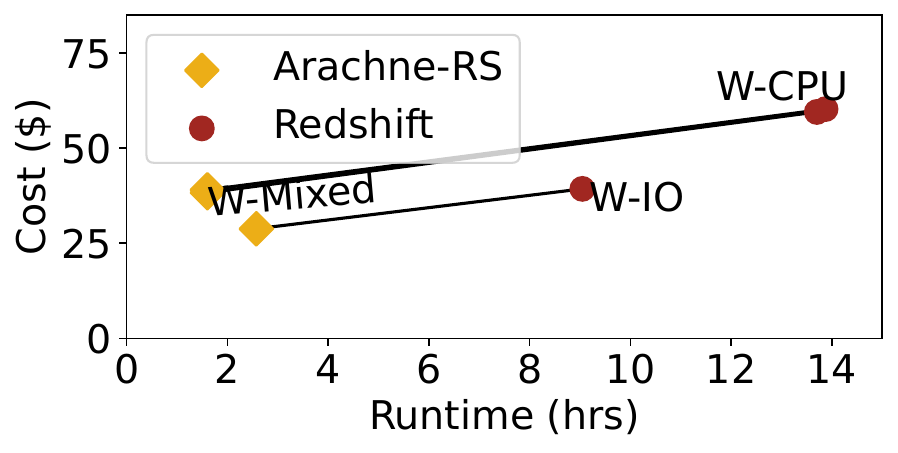}}\label{custom_aws}}%
    \subfloat[1TB G$\rightarrow$A4]%
    {{\includegraphics[width=0.5\columnwidth]{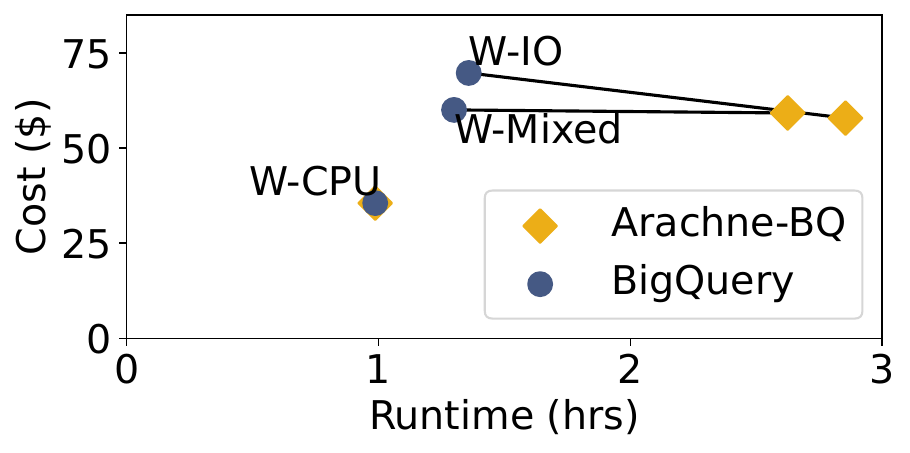}}\label{custom_gcp}}%
    \caption{(1TB A4$\rightarrow$G, G$\rightarrow$A4) Cost (USD) vs runtime
        (hours) for \textsc{W-CPU},	\textsc{W-IO}, and \textsc{Mixed} workload
    compositions}%
    \label{fig:custom}%
    \Description{There are two plots, one where data begins in AWS and one where data begins in GCP. When starting in AWS, Arachne's plans are both faster and cheaper. When starting in GCP, Arachne's plans for two workloads are cheaper but slower. For the third workload Arachne saves no money or time.}
\end{figure}

\mypar{Resource Balance Workload Overview} In Figure~\ref{fig:custom} we compare
the runtime (hours) on the x-axis and cost (USD) on the y-axis
of \sys{'s} execution plan to a baseline that executes
the workload in the starting backend. In
Figure~\ref{custom_aws}, data starts in Redshift. There are three red dots, one
for each workload, which represent the cost and runtime of executing that
workload in Redshift. 
The yellow dots represent the runtime and cost of executing that workload with \sys{}.
A line connects each yellow dot to its corresponding
red dot according to the workload. In Figure~\ref{custom_gcp}, data starts in
BigQuery (blue dots) instead of Redshift.

In 5 out of the 6 workloads \sys{} finds cheaper plans: all lines decrease from
the starting cloud baseline unless \sys{} has chosen the baseline execution
plan, and the degree of its reduction corresponds to monetary savings. 
For A4$\rightarrow$G in Figure~\ref{custom_aws}, \sys{} chooses multi-cloud
plans for all three workloads as there are enough CPU-bound queries to make
migration cost-effective. \sys{} saves 27\% on \textsc{W-IO} and 35\% on
\textsc{W-Mixed} and \textsc{W-CPU} over the Redshift baseline.  \sys{} saves
less money for \textsc{W-IO} because there are more IO-bound queries that favor
Redshift's per-second pricing model.
In G$\rightarrow$A4 in Figure~\ref{custom_gcp}, \sys{} executes
\textsc{W-CPU} entirely in BigQuery, so those two dots are
directly on top of each other. Since \textsc{W-Mixed} and \textsc{W-IO} contain
more IO-bound queries, \sys{} saves 1.35\% on \textsc{W-Mixed} and 17\% on
\textsc{W-IO} by migrating IO-bound queries to
Redshift. Because \textsc{W-IO} has \emph{more} IO-bound queries, the margin of
savings is larger for \textsc{W-IO} than for \textsc{W-Mixed}. 

Both \textsc{W-Mixed} and \textsc{W-CPU} include a very CPU-bound query, which
groups customers by spending history for recommendations. It runs in 6 hours and
costs \$25.84 in Redshift, while in BigQuery it runs in 3.5 minutes and costs \$1.
Other CPU-bound queries are similarly faster \emph{and} cheaper in
BigQuery. Consequently, in the A4$\rightarrow$G setup in
Figure~\ref{custom_aws}, baseline costs for \textsc{W-Mixed} and \textsc{W-CPU}
are similar, and \sys{} chooses similar multi-cloud plans for both workloads that
are faster \emph{and} cheaper than the Redshift baseline.


%

\begin{figure}%
    \centering
    \subfloat[A4$\rightarrow$G: Data Starts in Redshift]%
    {{\includegraphics[width=0.5\columnwidth]{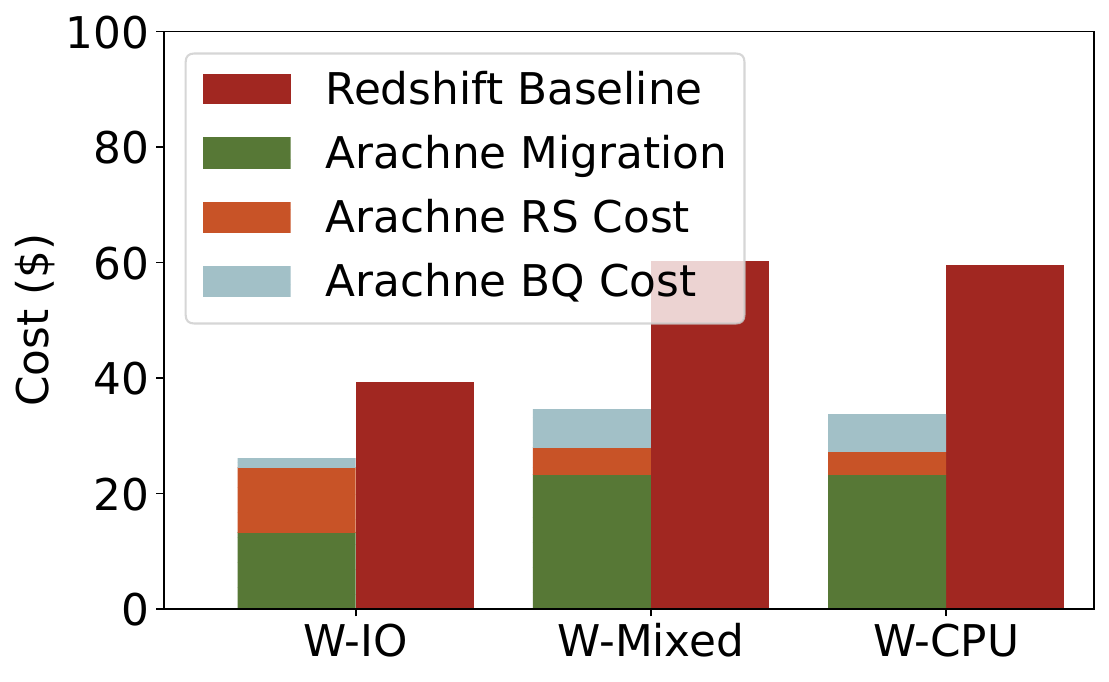}}\label{fig:custom_aws_breakdown}}%
    \subfloat[G$\rightarrow$A4: Data Starts in BigQuery]%
    {{\includegraphics[width=0.5\columnwidth]{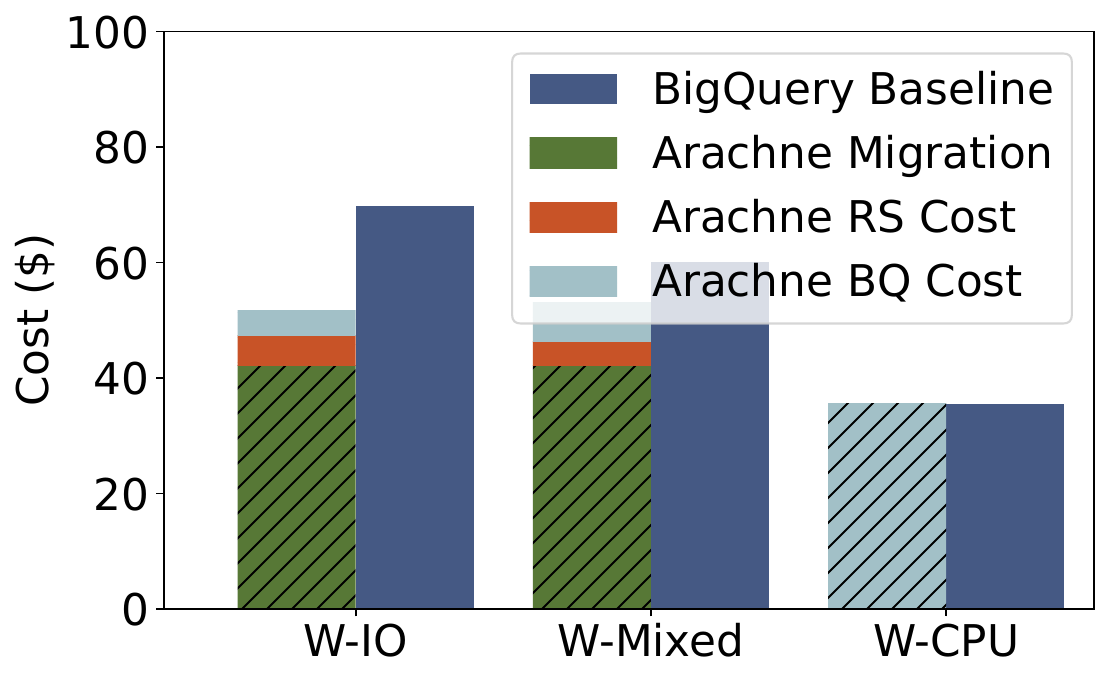}}\label{fig:custom_gcp_breakdown}}%
    \caption{(1TB) \sys{} migration costs, cost of queries moved, and cost
        of queries remaining versus baseline.
    }
    \label{fig:custom_breakdown}%
    \Description{Two plots are shown, each depicting three pairs of bars. One plot where data starts in AWS and one where data starts in GCP. For each pair of bars, the right bar is a solid color and represents the baseline. The left bar has diagonal lines through it and is broken up into three categories each with their own color: Arachne Migration, Arachne RS cost, and Arachne BQ cost.}
\end{figure}

%

\mypar{Resource Balance Workload Cost Breakdown} 
We divide costs for \sys{} into (1) migration costs, (2) the cost of queries which migrated, 
and (3) the cost of queries which remained in the source backend. 
In Figure~\ref{fig:custom_aws_breakdown} we breakdown costs for plans shown in
Figure~\ref{custom_aws} and in Figure~\ref{fig:custom_gcp_breakdown} we
breakdown costs for plans in Figure~\ref{custom_gcp}. Diagonal lines
are drawn through bars showing \sys{'s} plans. 

\begin{table}%
\caption{Inter-query plan-type by setup at 1TB and 2TB.}%
\centering
\resizebox{0.98\columnwidth}{!}{
\begin{tabular}{|c|c|c|c|c|c|}%
\hline%
Setup          & Arachne      & Multi & GCP & AWS & Total \\ \hline%
1TB, G$\rightarrow$A1, $\rightarrow$A4, $\rightarrow$A8        & \bf{23}      & 6     & 1   & 17  & 24    \\ \hline%
2TB, G$\rightarrow$A1        & \bf{22}      & 4     & 1   & 19  & 24    \\ \hline%
2TB, G$\rightarrow$A4        & \bf{22}      & 3     & 2   & 19  & 24    \\ \hline%
2TB, G$\rightarrow$A8        & \bf{22}      & 4     & 2   & 18  & 24    \\ \hline%
\end{tabular}%
}
\label{tbl:inter_distrib}%
\end{table}%

In Figure~\ref{fig:custom_gcp_breakdown},
\textsc{W-CPU} \sys{'s} bar is equal to the BigQuery baseline. If the
source pricing model is already favorable for a workload, \sys{} will keep all
data in the starting cloud. For all other plans in
Figure~\ref{fig:custom_breakdown},
multi-cloud savings are driven by the significantly lower execution cost for
queries that migrated to a favorable pricing model versus their baseline cost. In
Figure~\ref{fig:custom_breakdown}, the difference in query execution costs
is the baseline (right) bar minus the blue portion of \sys{'s} bar, which represents
the cost of queries remaining, presenting an enormous savings
opportunity. Exorbitant migration costs make up the majority of
\sys{'s} costs for multi-cloud plans. Egress is 90\% of all migration
costs--note that egress out of GCP is \$120/TB while egress out of AWS is
\$90/TB--loading data into Redshift is 5--8\% of migration costs, and the rest
is the cost of blob storage and data retrieval. 

\subsubsection{Read-Heavy Workloads}
\label{subsubsec:ds-derived}

Does \sys{} save money on skewed workloads? We first summarize the results before
zooming in on a few interesting workloads to understand the cost and runtime.


\mypar{Read-Heavy Overview} Table~\ref{tbl:inter_distrib} presents the outcomes
in setups \textsc{G$\rightarrow$A1}, \textsc{G$\rightarrow$A4},
\textsc{G$\rightarrow$A8}. The \textsc{Arachne} column indicates workloads where
\sys{} saves money over the BigQuery baseline. 
Across 144 workloads--48 workloads in 3 setups--only 6/144 remain in
BigQuery where the \sys{} saves no money. For workloads with
cheaper plans, in \textsc{Multi} plans some tables migrate 
to Redshift, and in \textsc{AWS} plans \emph{all} tables migrate to
Redshift. Since Read-Heavy workloads are IO-bound, the savings of moving
queries to Redshift compensate for migration costs. That even 6 of these
IO-skewed workloads remain in BigQuery demonstrates that egress costs
are a massive barrier to data movement.

At 2TB, we see that from G$\rightarrow$A1 to G$\rightarrow$A4 one \textsc{Multi}
plan flips to \textsc{GCP} and from G$\rightarrow$A4 to G$\rightarrow$A8 one
\textsc{AWS} plan flips to \textsc{Multi}. The 4$\times$ cost of
G$\rightarrow$A4 doesn't reduce runtime by 4$\times$, decreasing savings. 
If clusters are overprovisioned or underutilized, query savings will diminish
even with highly IO-bound workloads.

\sys{} saves up to 57.4\% on a single workload.
Of the 29 multi-cloud plans, most achieved 35\%--50\% savings. 
9 saved 2\%-8\%, while 1 saved less than 1\%. 
These plans save money by migrating queries to their most
beneficial pricing model.

\mypar{\textsc{Multi} Plan Analysis} We now focus on \textsc{Multi} plans which run
queries on both BigQuery and Redshift to show the
opportunities of combining \emph{price-per-byte} and \emph{price-per-compute}
pricing models.

\begin{figure}%
    \centering
    \subfloat[1TB G$\rightarrow$A1]{{\includegraphics[width=0.5\columnwidth]{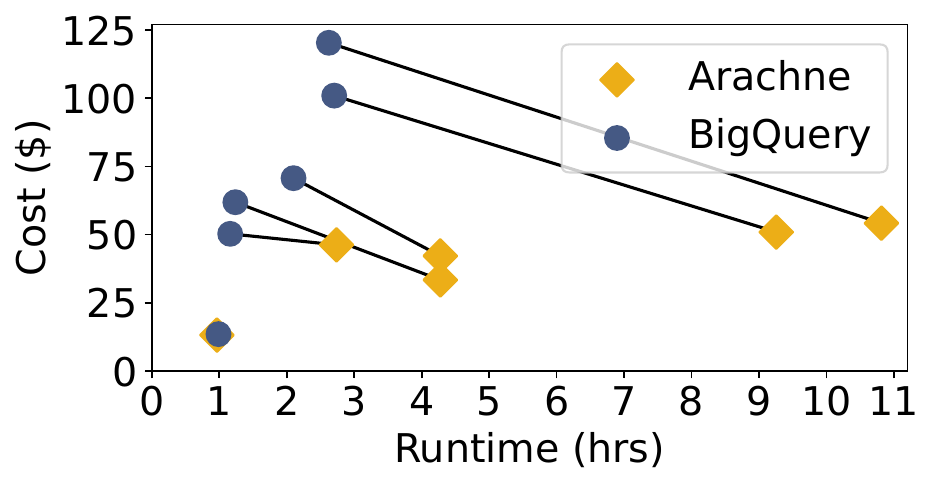}}\label{sf1000_ga1}}%
    \subfloat[2TB G$\rightarrow$A1]{{\includegraphics[width=0.5\columnwidth]{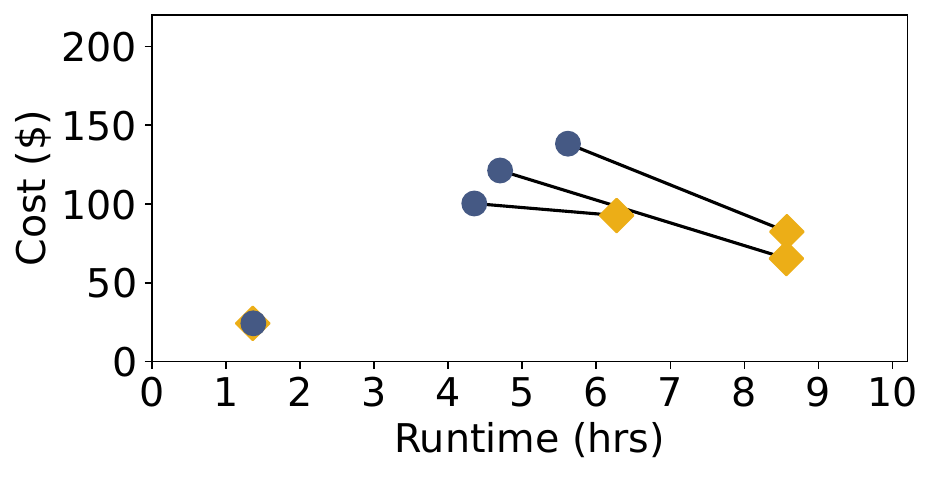}}\label{sf2000_ga1}} \\ 
    \subfloat[1TB G$\rightarrow$A4]{{\includegraphics[width=0.5\columnwidth]{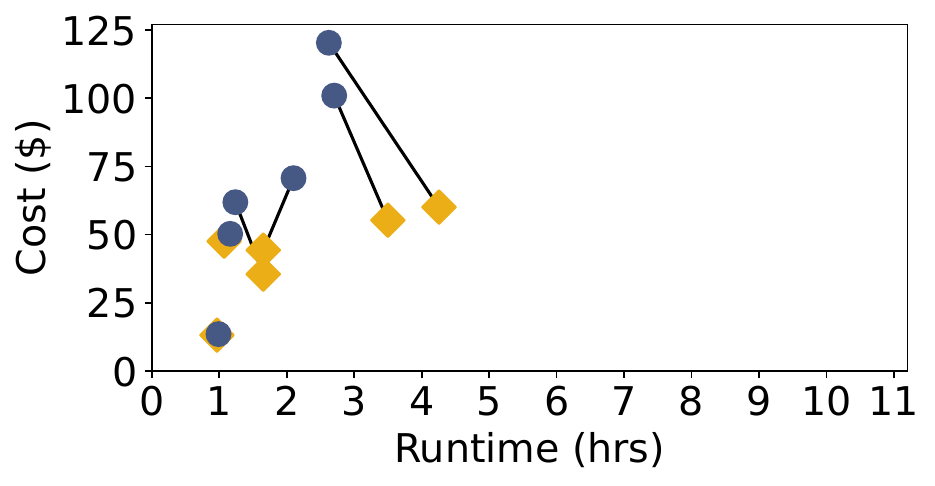}}\label{sf1000_ga4}}%
    \subfloat[2TB G$\rightarrow$A4]{{\includegraphics[width=0.5\columnwidth]{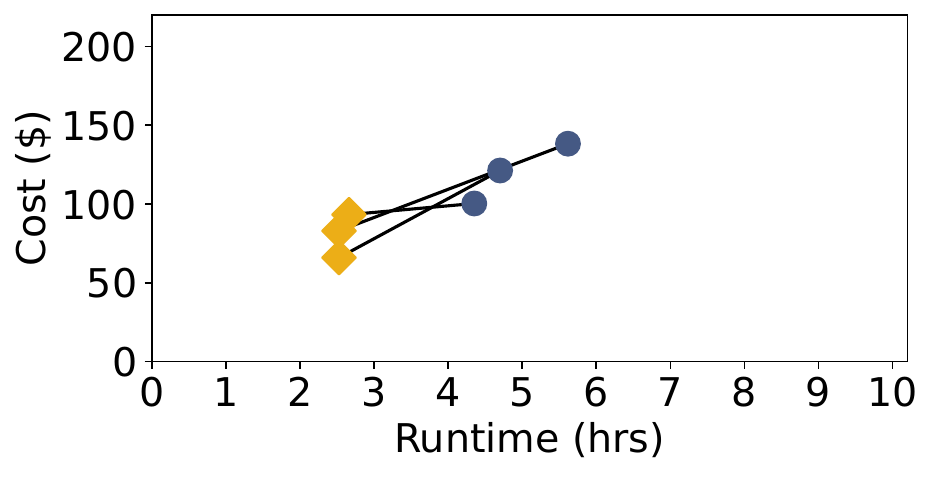}}\label{sf2000_g4}} \\
    \subfloat[1TB G$\rightarrow$A8]{{\includegraphics[width=0.5\columnwidth]{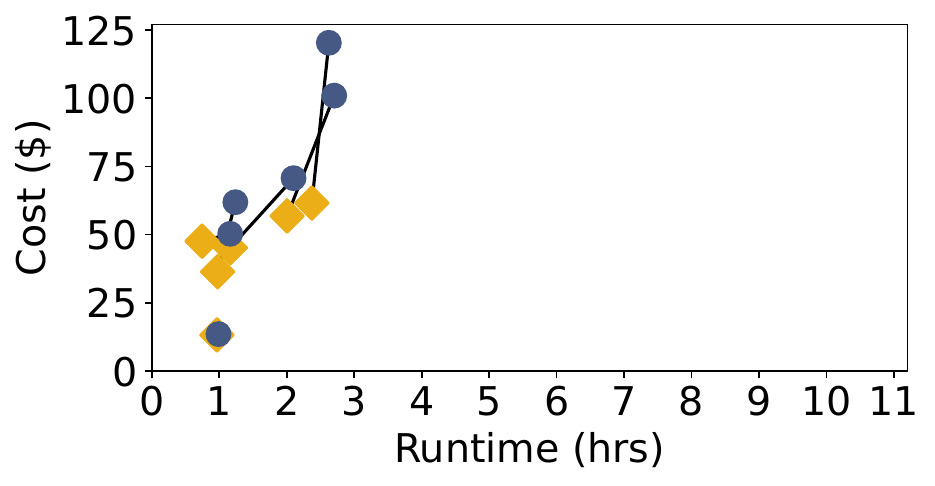}}\label{sf1000_ga8}}%
    \subfloat[2TB G$\rightarrow$A8]{{\includegraphics[width=0.5\columnwidth]{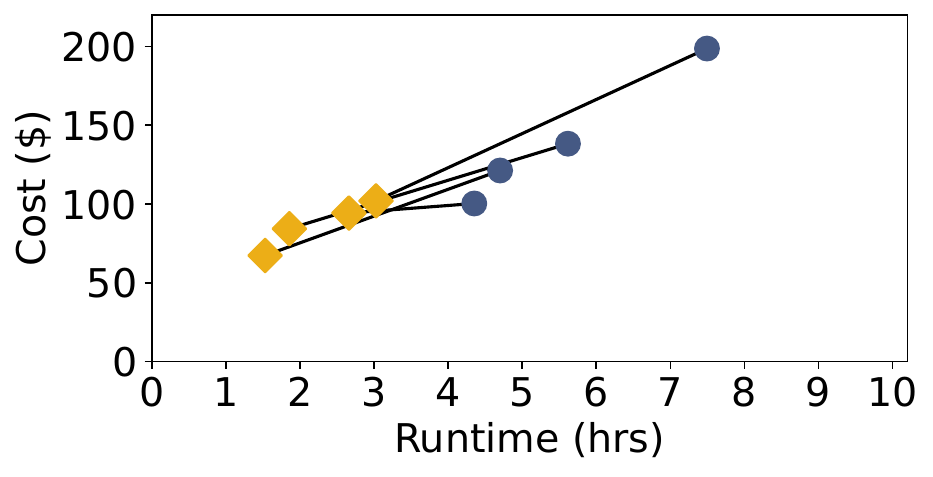}}\label{sf2000_ga8}}%
    \caption{Cost (USD) vs runtime (hours) for 1--2TB datasets with \textsc{Multi}
    plans over Redshift and BigQuery.}%
    \label{fig:sf1000_runtime}%
    \Description{Six plots are shown in two sets of 3. Each set shows three different setups, one using a 1 node Redshift cluster, one using a 4 node Redshift cluster, and one using an 8 node Redshift cluster. The first set is 1TB dataset and the second set is a 2TB dataset. There are yellow dots and corresponding blue dots that are connected to a yellow dot by a line. For 1TB and 1 node, all Arachne plans are slower and cheaper, except for one which is negligibly cheaper and slower. For 1TB and 4 nodes, 2 workloads are cheaper and faster, 2 are cheaper and slower, and 2 have marginal differences. For 1TB and 8 nodes, all workloads are both cheaper and faster. For 2TB and 4 nodes the workloads are cheaper and nearly twice as fast. At 2TB and 8 nodes many workloads are nearly 3 times faster.}
\end{figure}

As in Figure~\ref{fig:custom}, Figure~\ref{fig:sf1000_runtime} plots workload
cost versus runtime for \textsc{Multi} plans. Blue dots represent BigQuery
baseline plans while yellow dots represent \sys{} plans. Dots corresponding to
the same workload are connected with a line. These plans achieve up to 54\%
savings and over 35\% for most workloads. 2 workloads at 1TB and 1 workload at
2TB save between 2 and 8\%. 1 workload at 2TB G$\rightarrow$A1 saves less than
1\%, as \sys{} migrates very few queries and tables which yield marginal
savings.

In Figure~\ref{fig:sf1000_runtime}, the BigQuery baseline is faster than
\sys{'s} plan because \textsc{G$\rightarrow$A1} does not exploit all the parallelism
available in the workload. At \textsc{G$\rightarrow$A4} we see that \sys{'s} plan is cheaper
and closer in runtime to the baseline, and is both cheaper and faster in
\textsc{G$\rightarrow$A8}. In larger clusters, loading times decrease, and Redshift completes
the workloads faster (and cheaper) than BigQuery. At 2TB, the trends are similar
to 1TB except that now \sys{} is both cheaper and faster than BigQuery
even in the \textsc{G$\rightarrow$A4} case. 
\begin{figure}
    \centering
    \includegraphics[width=0.9\columnwidth]{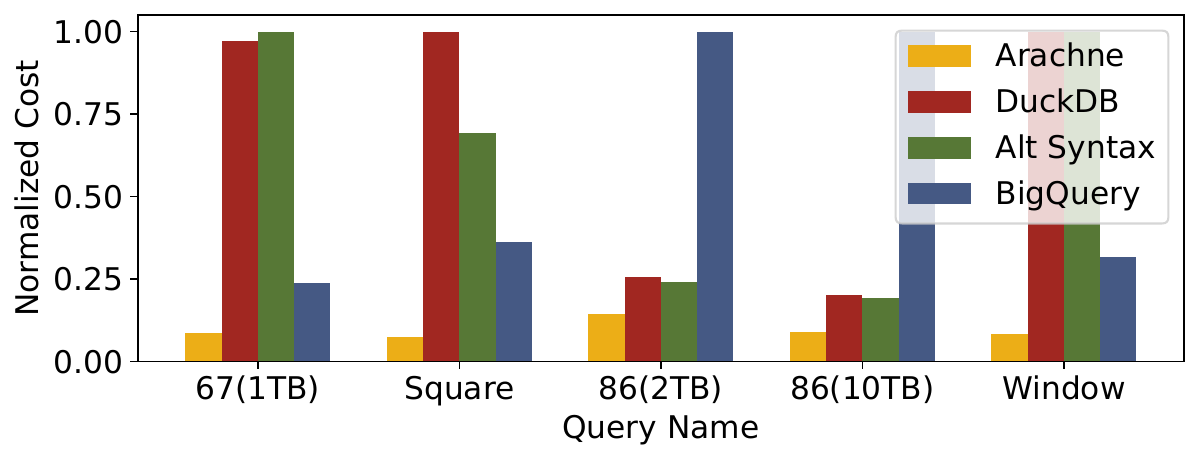}
    \caption{Query costs normalized to most expensive plan for
        \sys{'s} intra-query plan vs BigQuery, DuckDB, and
        DuckDB with \sys{}-produced text (Alt Syntax).}
    \label{fig:gcp_hybrids}
    \Description{Shows 5 sets of 4 bars. Plots normalized cost on the y-axis. Each set represents a different query. Each of the 4 bars represents a different baseline. The yellow bar representing Arachne is the lowest in all 5 cases. Which baseline is the most expensive changes between queries. The yellow bar is 30-35 times cheaper than the most expensive baseline in all cases and at least 2-3 times cheaper than the next cheapest baseline.}
\end{figure}

\mypar{BigQuery Internal Tables} Loading data into BigQuery is free but
significantly increases runtime; loading 1TB took 12 minutes whereas creating
external tables took only 20 seconds. Data is stored in a closed format and only
accessible via their SQL interface, which incurs query
processing costs, or BigQuery's Storage Read API~\cite{bq_storage_price}; both are much more expensive than the blob storage API costs. 

For queries over internal tables, BigQuery charges once for each table scanned,
even if the table is scanned multiple times in the query. For queries over
external tables, BigQuery charges for \emph{each} table scan operator, even if
multiple operators scan the same table. So the same query will scan fewer bytes
and cost less when data is stored internally. We run the inter-query algorithm
on G$\rightarrow$A1, G$\rightarrow$A4, and G$\rightarrow$A8 with data stored
internally. There are 3-5 multi-cloud workloads in each setup saving 3--20\% and
2 with negligible savings, similar to the external case, because of high BigQuery
prices and the large IO-bound savings in the Read-Heavy workloads. These savings
margins are smaller due to the fewer bytes billed.

\subsubsection{Extending \sys{} with IaaS+DuckDB}
\label{subsubsec:iaas}

We now explore the cost differences between PaaS
(the databases evaluated above) and a new execution backend that deploys DuckDB
on a GCP VM, where data starts, to avoid egress costs.  The VM costs
\$1.49/hour with 16
vCPU, 190GB RAM, and 1TB disk to run memory-intensive queries. 

This VM did not have enough memory to run all queries. To concentrate on
studying the effect of IaaS, we edited the queries slightly, replacing WITH
clauses with CREATE TABLE AS clauses so that intermediate tables are offloaded
to disk. While this may increase query runtime it allowed the query to complete
so we could proceed with our study. We only consider queries which execute in
all execution backends.

In this setup, \sys{} does not migrate any queries to Redshift, as IaaS lowered
query costs enough that migration is not worthwhile. However, \sys{} still
achieves up to 55\% savings over the pay-per-byte BigQuery (PaaS) baseline by
utilizing cheaper, pay-per-compute IaaS which dramatically reduces costs for the
IO-bound workloads. Hence, \sys{} can identify opportunities and achieve
significant savings with a transparent deployment of DuckDB on IaaS, all without
separate user setup, deployment, or maintenance. 

\subsubsection{Summary} 

\sys{} successfully exploits the
inter-query algorithm (\textbf{O1}), even in highly skewed workloads if the
source pricing model is ineffective for it. \sys{} chooses
multi-cloud plans saving 35\%--56\% in most cases
by using multiple pricing models. 
\begin{figure*}%
    \centering
    \subfloat[A4$\rightarrow$G BigQuery Price]{{\includegraphics[width=0.25\textwidth]{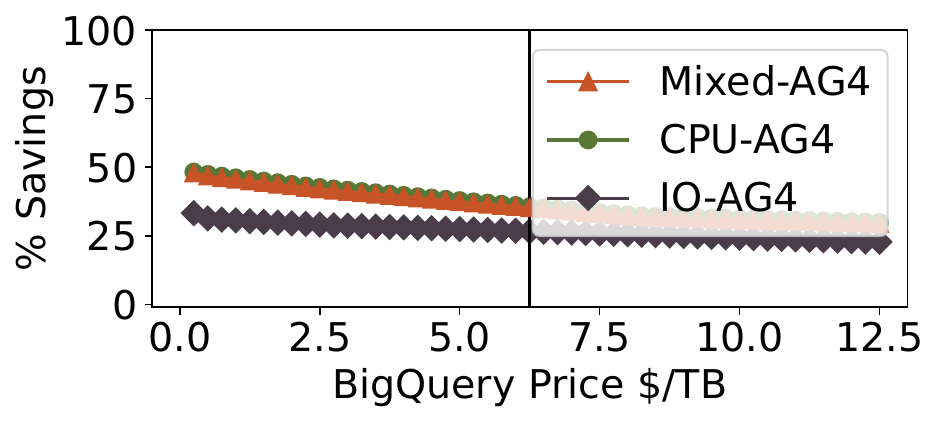}}\label{fig:aws_bq_percent}}%
    \subfloat[A4$\rightarrow$G Egress Price]{{\includegraphics[width=0.25\textwidth]{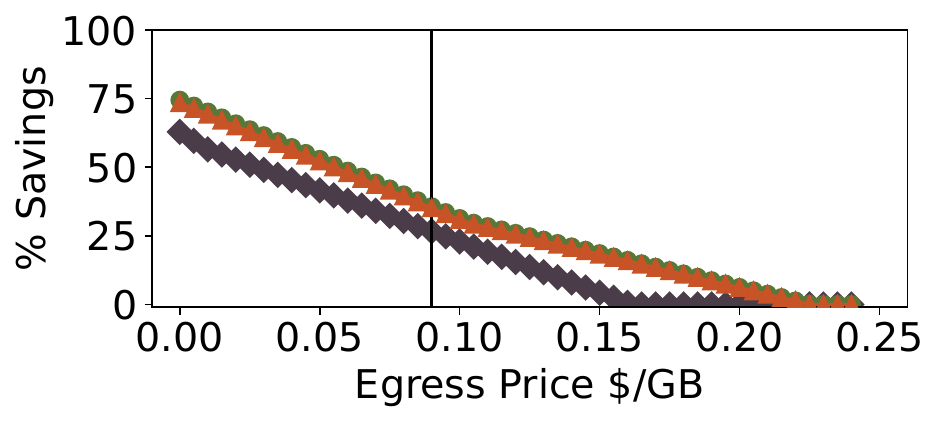}}\label{fig:aws_e_percent}}%
    \subfloat[G$\rightarrow$A4 BigQuery Price]{{\includegraphics[width=0.25\textwidth]{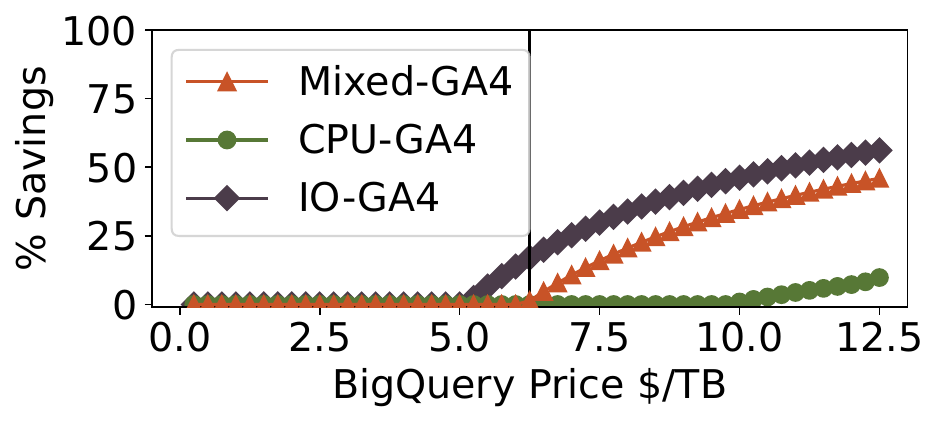}}\label{fig:gcp_bq_percent}}%
    \subfloat[G$\rightarrow$A4 Egress Price]{{\includegraphics[width=0.25\textwidth]{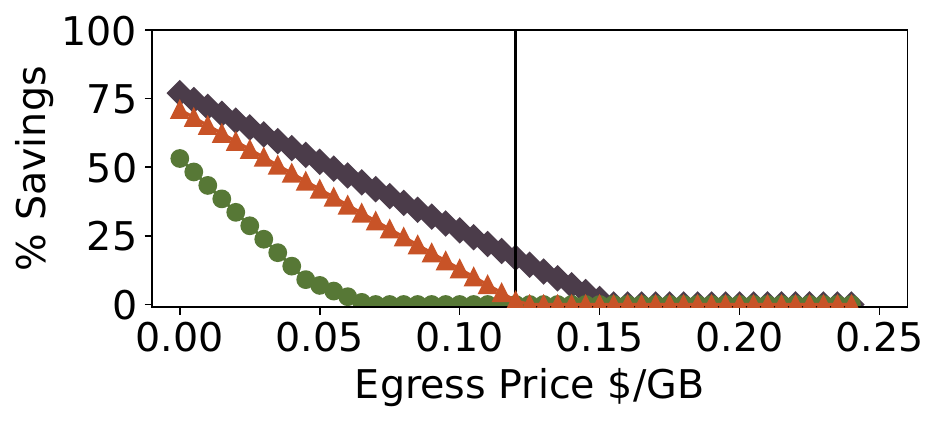}}\label{fig:gcp_e_percent}}%
    \caption{(G$\rightarrow$A4, A4$\rightarrow$G 1TB) \% savings for inter-query plans vs. BigQuery or egress prices on Resource Balance Workloads}%
    \label{fig:custom_percents}%
    \Description{Four plots showing relative savings versus varying either egress price or BigQuery price in 2 different setups. As egress price increases eventually all savings go to zero. When starting in BigQuery, very low BigQuery prices cause no savings but as prices increase there is a point where all workloads save money via inter-query plans. WHen data starts in AWS, there is smaller changes with savings as BigQuery price increases, and it is always above zero.}
\end{figure*}

\begin{figure}%
    \centering
    \subfloat[1TB Varying BigQuery Price]{{\includegraphics[width=0.25\textwidth]{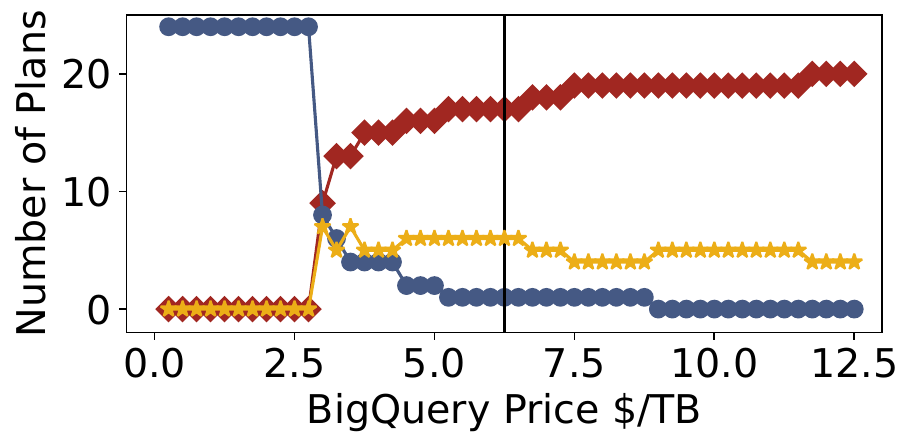}}\label{fig:sf1000_bq}}%
    \subfloat[1TB Varying Egress Price]{{\includegraphics[width=0.25\textwidth]{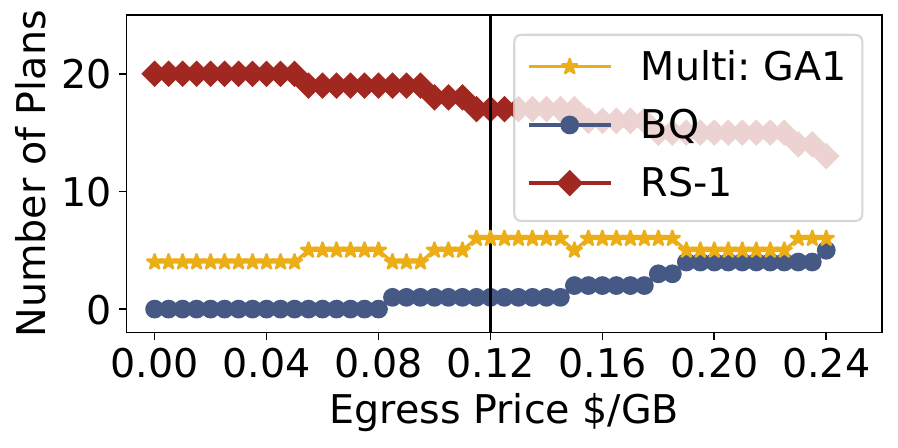}}\label{fig:sf1000_e}}%
    \caption{(G$\rightarrow$A1 1TB) Inter-query results varying either BigQuery
    or egress price on Read-Heavy Workloads }%
    \label{fig:what_if}%
    \Description{As BigQuery price increases, the number of BQ-only plans drops precipitously while the number of AWS-only plans increase. The number of multi-plans is nonzero and increases for a while but then settles and fluctuates slightly. As egress price increases, there are more BQ-only plans and fewer AWS-only plans while the number of multi-cloud plans varies somewhat but stays relatively consistent.}
\end{figure}


\begin{table}
\centering
\caption{Absolute baseline and Arachne intra-query costs. 
}
\begin{tabular}{|c|c|c|c|c|}
\hline
Query      & Arachne    & BigQuery  & DuckDB    & Alt Syntax \\ \hline
67         & \$1.83     & \$4.9981  & \$20.4027 & \$21.0109  \\ \hline
Square     & \$0.005507 & \$0.0156  & \$0.07321 & \$0.05069  \\ \hline
86 (2TB)   & \$0.089574 & \$0.62853 & \$0.1605  & \$0.15144  \\ \hline
86 (10TB)  & \$0.278728 & \$3.142   & \$0.63195 & \$0.60877  \\ \hline
Window     & \$0.311999 & \$1.1791  & \$3.7159  & \$3.7159   \\ \hline
\end{tabular}
\label{tbl:hybrid_abs}
\end{table}

\subsection{RQ2. Intra-Query Processing}
In this section, we explore opportunity \textbf{O2} via the intra-query
algorithm. 
These opportunities are significant but occur less often than \textbf{O1},
so we report results for
five queries that produced a cheaper intra-query plan and analyze
the characteristics of these queries.

\mypar{Queries} Query 67 and \textsc{window} are run on a
1TB TPC-DS dataset and query 86 is run on a 2TB and 10TB TPC-DS dataset.
The \textsc{window} query performs several joins and group-bys and
executes a complex window operation on the result. The
\textsc{square} query, run on 100GB LDBC-SNB dataset, finds 
squares in social media graphs, \eg a path from person A to B to C to D and back to A. 

\mypar{Experimental Setup} Data starts in BigQuery, and we consider intra-query
plans between BigQuery (pay-per-byte) and DuckDB (pay-per-compute) on a GCP VM.
Expensive egress fees restrict data movement and eliminate
    intra-query opportunities across multiple
clouds, so we consider GCP-only intra-query plans.
We pay profiling
costs to copy data to the VM and execute queries in BigQuery and
DuckDB, gathering cost, runtime, and operator cardinality.
\sys{} converts its internal query plan into SQL to execute subqueries, but we
observe that alternate SQL texts for the same logical query can cause the optimizer to
choose different physical plans, affecting runtime and cost. To isolate this
factor, we consider a third baseline, called \textbf{Alt Syntax}, which is the
cost of executing the initial query rewritten by \sys{} in DuckDB. 

\mypar{Results} Figure~\ref{fig:gcp_hybrids} shows the costs for \sys{'s}
intra-query plan and the baselines normalized to the most expensive baseline for
each query. \sys{'s} plans save 2--5$\times$ compared to the next cheapest
baseline and orders of magnitude compared to the most expensive baseline,
showing the cost saving potential of \textbf{O2}. We normalize values to
emphasize the relative savings as the absolute savings are small because
out-of-memory errors on larger datasets prevented us from using longer-running
queries. We show the absolute numbers in Table~\ref{tbl:hybrid_abs} and note
that relative costs better represent total savings, as the total savings are
query savings multiplied by the number of times a query executes. While
intra-query plans sometimes run slower than the fastest baseline (shown in
Table~\ref{tbl:hybrid_run}), this potential slowdown is tolerable if the
query is run in a latency-insensitive, periodic workload such as a nightly
analytics workload.

\begin{table}
\caption{Runtime (seconds) for baselines and Arachne plans.}
\centering
\begin{tabular}{|c|c|c|c|c|}
\hline
Query      & Arachne    & BigQuery  & DuckDB    & Alt Syntax \\ \hline
67         & 4059.96 & 555.333 & 50655.30 & 51107.595  \\ \hline 
Square     & 188.226 & 14.569  & 168.727  & 113.961    \\ \hline
86 (2TB)   & 171.051 & 206.045 & 381.271  & 359.023    \\ \hline
86 (10TB)  & 580.898 & 423.063 & 1527.825 & 1471.458   \\ \hline
Window     & 624.970 & 82.155  & 9038.641 & 8954.334   \\ \hline
\end{tabular}
\label{tbl:hybrid_run}
\end{table}

\mypar{Common Characteristics} These queries first join many tables (IO-bound)
followed by a window or self-join (CPU-bound).  Queries with these stages are
good candidates for the intra-query algorithm.


\mypar{Summary} There are fewer situations where \textbf{O2} saves money versus
\textbf{O1}, but when opportunities exist, the relative savings are significant,
especially for queries with the structure discussed above. Profiling costs
for 3/5 queries are earned back in under 25 iterations. Query 67 and
\textsc{window} are earned back in 28 and 46 iterations and cost \$85.68 and
\$40.18. The savings achieved are significant and compensate for
incurred profiling costs. Re-profiling may be
    required more frequently for \textbf{O2} than \textbf{O1}, 
increasing costs. However, the sizable savings margin can quickly earn back that
up-front cost.


\subsection{RQ3. Simulating Different Cloud Costs}
\label{subsec:what-if}


So far, the results shown assume cloud vendor prices as of February'24. In this
section, we use profiled inputs (discussed in Section~\ref{subsec:profiler})
that are not affected by cloud vendor
prices and simulate cloud prices by varying the price inputs to
the inter-query algorithm. We vary the price-per-byte (BigQuery
price) and egress price from the source execution backend and run the
inter-query algorithm on the Read-Heavy (RH) workloads and Resource Balance Workloads
(RBW) to see how varying prices impacts savings. Vertical lines 
indicate the current price. We use the plan types as in
Section~\ref{subsec:inter-query}--\textsc{GCP}, \textsc{AWS}, or \textsc{Multi}.
For RBW, we show percent savings (Figure~\ref{fig:custom_percents}) versus the
price being varied in A4$\rightarrow$G and G$\rightarrow$A4.
Figure~\ref{fig:what_if} shows plan types for RH in G$\rightarrow$A1 at 1TB;
trends are similar in G$\rightarrow$A4 and G$\rightarrow$A8 and at 2TB. The main takeaways
are:

\begin{myitemize}
\item Inter-query savings (\textbf{O1}) are robust to changes in prices even for
    heavily IO-bound workloads. 
\item Reducing BigQuery price by 40\% to \$3.75/TB keeps most RH plans
    in BigQuery. At 2TB the necessary price reduction increases because potential savings
    grow as dataset size grows.
\item For RBW, in A4$\rightarrow$G BigQuery price does not impact plan
    type but slightly reduces savings. In G$\rightarrow$A4, reducing 
    prices by 20\% to \$5/TB keeps all plans in BigQuery. 
\item At low egress prices, \textsc{Multi} plans are the cheapest option for
    4/24 RH workloads in 1--2TB and all RBWs in A4$\rightarrow$G.
    Even if cloud vendors lower financial barriers to data movement, 
    money-saving, inter-query opportunities (\textbf{O1}) still exist.
\item High egress prices lock-in all RBW plans to their starting cloud. For
    RH, multi-cloud plans still exist, so savings achievable by
    using multiple pricing models outpace egress costs. 
\end{myitemize}

\begin{figure}%
    \centering
    \subfloat[1TB G$\rightarrow$A4 Varying BigQuery Price]%
    {{\includegraphics[width=0.5\columnwidth]{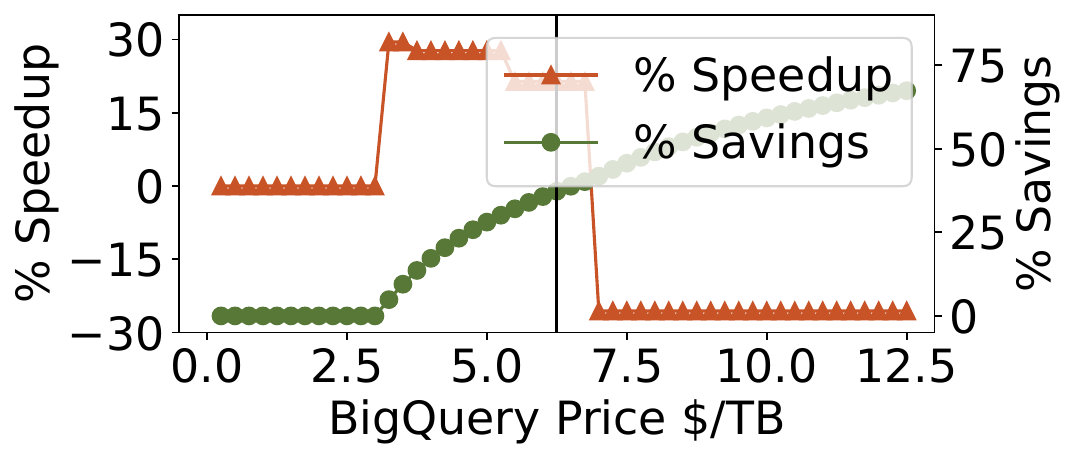}}\label{ws_bq}}%
    \subfloat[1TB G$\rightarrow$A4 Varying Egress Price]%
    {{\includegraphics[width=0.5\columnwidth]{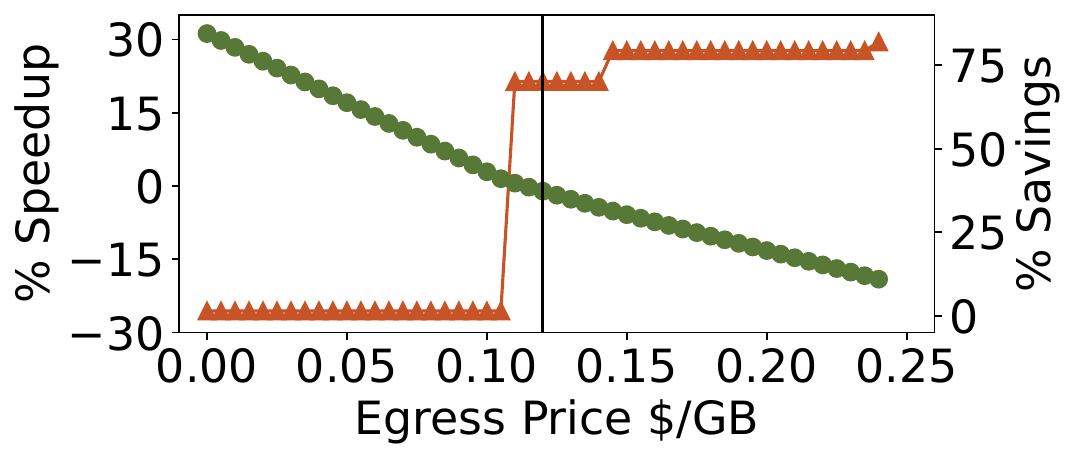}}\label{ws_e}}%
    \caption{(1TB G$\rightarrow$A4) \% Savings and speedup of \sys{} 
        for Read-Heavy 22 vs. BigQuery and egress price.
    }%
    \label{fig:ws_what_if}%
    \Description{Shows two plots for 1TB and starting in GCP and going to an A4 setup. As BigQuery price increases the savings go from 0 to about 75 percent in a curve that is concave down. Runtime is the same when savings are 0, is faster by about 30 percent for lower amounts of savings, and then is 30 percent slower for the plans with greater savings at higher BigQuery prices. When increasing egress price savings start at about 75 percent and decrease, roughly concave up, down to close to 0. The larger savings plans are about 30 percent slower but then get faster as savings decrease, going up to between 15 and 30 percent faster.}
\end{figure}

\mypar{Runtime vs. Cost Tradeoffs}
To observe how cloud prices impact cost and runtime tradeoffs, we zoom in on
\textbf{Read-Heavy 22} at G$\rightarrow$A4 1TB and show the percent savings 
and percent speedup over the baseline vs. BigQuery and egress prices
in Figure~\ref{fig:ws_what_if}. 
Negative percent speedup indicates that \sys{'s} plan is slower.

A small increase in BigQuery price from \$6.25/TB to \$7/TB causes \sys{} to
migrate more tables in a multi-cloud plan, which runs longer than the baseline
but achieves greater savings. 
A slight decrease of egress cost to \$0.105/GB
from \$0.12/GB yields a similar result for the same reasons, as migrating more
tables increases savings but also runtime. At many other prices the \sys{'s}
plan is both cheaper and faster. Figure~\ref{fig:ws_what_if} illustrates how
the specific tradeoff of runtime and cost is impacted by cloud vendor prices.

\mypar{Conclusions} Overall, we see that our results are not brittle to price
changes: some queries are simply cheaper in different pricing models and 
prices dictate how large savings are and how large barriers to
migration are. More importantly, they show the power platforms have to lock in
workloads by adjusting prices slightly; this anti-competitive restriction should
be concerning to all of us.

\subsection{RQ4: Profiling Cost Microbenchmarks}
\label{subsec:micro}
Gathering inter-query inputs (Section~\ref{subsec:profiler}) incurs significant
\emph{profiling} costs. 
We show how stale profiles impact
savings (Section~\ref{subsec:time-series}), how profiling over samples lowers
profiling costs (Section~\ref{subsubsec:sampling}), and how noisy runtime
estimates impact savings (Section~\ref{subsec:estimation})

\subsubsection{Impact of Re-profiling}
\label{subsec:time-series}

We first study the savings impact of using stale
    profiles as data changes. We create 7 datasets with sizes from
    100--1200GB with the official TPC-DS generator. A TPC-DS dataset
    reflects a database at a moment in time for a retail supplier.
    While tables tracking sales and returns only grow with overall data size,
    other tables tracking inventory or customers grow and shrink as overall data
    size increases, per the TPC-DS
    specification~\cite{tpcds_spec}.

\begin{figure}%
    \centering
    {\includegraphics[width=\columnwidth]{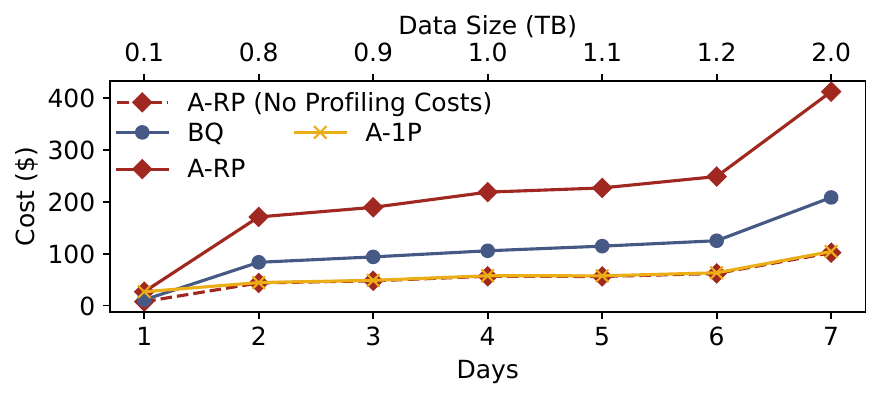}}%
    \caption{
        (G$\rightarrow$A4) Cost (USD) vs days and data size (TB)
    }%
    \label{fig:time-series}%
    \Description{One plot showing the cost of four different execution strategies versus time in days. Each day has a corresponding data size shown on the upper x-axis. Days are numbered 1 to 7. Day 1 has 0.1TB, Day 2 has 0.8TB, Day 3 has 0.9TB, Day 4 has 1.0TB, Day 5 has 1.1TB, Day 6 has 1.2TB, and Day 7 has 2TB. All four points are clustered together with similar costs on Day 1. The red line for A-RP is the highest across all days, and from day 2 to 7 is significantly higher than the rest. The blue line representing the BigQuery baseline is always cheaper than that but always second most expensive. The yellow line is marginally more expensive than the dotted red line, but is consistently the cheapest of the four.}
\end{figure}

We let each data size be the state on a given day over 7 days. In Figure~\ref{fig:time-series} we
compare four execution strategies 
in G$\rightarrow$A4 as data changes.
We compare the BigQuery baseline (BQ), 
against two \sys{} strategies: profiling only on day 1
(A-1P) and using that profile 
for days 2--7; and profiling as soon as data changes (the solid red
line, A-RP). A-1P and A-RP include profiling costs.
Finally, we show the cost of A-RP without
profiling costs as the dotted red line. 

We compare these strategies on Read-Heavy 2 which showed the largest gap
between A-1P and A-RP.
While A-RP is cheaper over time, it is
at most 2\% cheaper than A-1P. Daily
re-profiling costs make A-RP far more expensive than BQ. A-1P quickly
compensates for profiling costs and saves significantly over BQ. A-1P's stale
profile still captures which queries are cheaper in which backend,
so small errors in the profile do not appreciably diminish savings.

\subsubsection{Sampling}
\label{subsubsec:sampling}


We now show how sampling reduces profiling costs with low error.
While estimating runtime from samples is difficult for some
queries~\cite{surajit_sampling, liang_combining_2021, yang_deep_2019}, most
profiling cost are from \emph{pay-per-byte} pricing models, where runtime does
not affect cost. 

We show cost and estimation error for samples of 15, 25, 50, and
100\% of data in Table~\ref{tbl:profiling_sample}.
When profiling over all data, 20/24 workloads earn back profiling costs in
4 iterations. Small samples estimate the inputs well and in most cases
lower the number of needed iterations to 1--2.
Read-Heavy 7 chooses a GCP-only plan so achieves no savings. Read-Heavy 17
only achieves marginal savings and needs many iterations, though sampling
lowers the net cost of profiling. We do not claim that sampling is the
best approach, only that it cheapens profiling. More sophisticated
approaches, \eg using parameterized cost models to sample non-linear
operators~\cite{li_abstract_2019} can further reduce error, which we leave for
future work.



\subsubsection{Profiling versus Runtime Estimation}
\label{subsec:estimation}

We estimate query runtimes in Redshift by training a Kernel
    Canonical Correlation Analysis (KCCA) model, as proposed by Ganapathi et.
    al.~\cite{estimation} in 2009\footnote{The authors of the original
    paper could not provide the materials to reproduce their work; we replicated
their model effort to the best of our ability}. KCCA finds correlated clusters
of training features and labels to make predictions.  However, hardware advances
over 15 years mean that queries run much faster, so most training points 
are clustered together despite the size and diversity of the
training set, lowering the reproduced model's accuracy. 
Nonetheless, we created 2842 training queries
as the original 3102 queries used were not available after reaching out to the
authors. We used a 1GB TPC-DS dataset 
and also ran some queries on 100GB and 1TB datasets to get
a broader range of runtimes. We make a significant effort to replicate the setup
and create a runtime estimation method for SQL queries.

Inter-query plans using runtime estimates are 66\% more expensive
than plans using profiles on \textsc{W-Mixed} in the A1$\rightarrow$G setup and
13\% more expensive in G$\rightarrow$A1 when they cost the same as the baseline.
Noisy estimates result in \sys{} missing valuable savings
opportunities and greatly diminish savings margins, illustrating the
detrimental impact of estimation on inter-query savings.


\section{Related Work}
\label{sec:relatedwork}

\mypar{Other Cloud Databases} Many other cloud and third-party databases scan
cloud storage like Snowflake, Azure Synapse, Trino, Apache Hive, Amazon Athena,
and SparkSQL \cite{snowflake,synapse,polaris,presto,hive,athena,sparksql}. These
databases use per-second billing (Presto, Hive, SparkSQL), per-byte billing
(Athena), or some combination of the two.  While we could have used other
systems in our evaluation, Google BigQuery and AWS Redshift effectively
represent both pricing models across clouds, enabling us to evaluate
opportunities \textbf{O1}--\textbf{O2}.


\mypar{Cloud Cost Savings}
Prior work on money savings focuses on scheduling algorithms
\cite{wu_workflow_2015}, exploring cost sources in different execution backends
\cite{tan_choosing_2019}, or using S3
Select~\cite{s3_select} to speed up queries and lower costs~\cite{pushdowndb}.
Leis and Kuschewski model per-second costs for cloud workloads
\cite{leis_towards_2021}. Recent work 
has also focused on achieving savings using spot
instances~\cite{spot_sky},
minimizing network egress prices~\cite{cloudcast}, and finding cheaper
configurations for cloud deployments~\cite{bangoptimizing}.
To the best of our knowledge, \sys{} is the first
effort to systematically explore savings opportunities for analytical
queries by using multiple databases with different pricing models.

\begin{table}[]
\centering
\resizebox{\columnwidth}{!}{
\begin{tabular}{|c|ccc|ccc|ccc|ccc|}
\hline
Sample \%  & \multicolumn{3}{c|}{15}         & \multicolumn{3}{c|}{25}         & \multicolumn{3}{c|}{50}          & \multicolumn{3}{c|}{100}          \\ \hline
Dataset & \multicolumn{1}{c|}{Cost} & \multicolumn{1}{c|}{Iter} & \multicolumn{1}{c|}{Error} & \multicolumn{1}{c|}{Cost} & \multicolumn{1}{c|}{Iter} & \multicolumn{1}{c|}{Error} & \multicolumn{1}{c|}{Cost} & \multicolumn{1}{c|}{Iter} & \multicolumn{1}{c|}{Error} & \multicolumn{1}{c|}{Cost} & \multicolumn{1}{c|}{Iter} & \multicolumn{1}{c|}{Error} \\ \hline
Read-Heavy 0 & \multicolumn{1}{c|}{30.39} & \multicolumn{1}{c|}{1} & 0.02 & \multicolumn{1}{c|}{49.33} & \multicolumn{1}{c|}{1} & 0.03 & \multicolumn{1}{c|}{94.2} & \multicolumn{1}{c|}{2} & 0.03 & \multicolumn{1}{c|}{177.19} & \multicolumn{1}{c|}{3} & 0.0 \\ \hline
Read-Heavy 1 & \multicolumn{1}{c|}{30.04} & \multicolumn{1}{c|}{1} & 0.02 & \multicolumn{1}{c|}{48.71} & \multicolumn{1}{c|}{1} & 0.03 & \multicolumn{1}{c|}{92.97} & \multicolumn{1}{c|}{2} & 0.03 & \multicolumn{1}{c|}{174.93} & \multicolumn{1}{c|}{3} & 0.0 \\ \hline
Read-Heavy 2 & \multicolumn{1}{c|}{27.41} & \multicolumn{1}{c|}{1} & 0.02 & \multicolumn{1}{c|}{44.49} & \multicolumn{1}{c|}{1} & 0.03 & \multicolumn{1}{c|}{84.54} & \multicolumn{1}{c|}{2} & 0.04 & \multicolumn{1}{c|}{157.69} & \multicolumn{1}{c|}{3} & 0.0 \\ \hline
Read-Heavy 3 & \multicolumn{1}{c|}{17.54} & \multicolumn{1}{c|}{1} & 0.03 & \multicolumn{1}{c|}{28.17} & \multicolumn{1}{c|}{1} & 0.04 & \multicolumn{1}{c|}{53.45} & \multicolumn{1}{c|}{2} & 0.05 & \multicolumn{1}{c|}{97.78} & \multicolumn{1}{c|}{4} & 0.0 \\ \hline
Read-Heavy 4 & \multicolumn{1}{c|}{25.6} & \multicolumn{1}{c|}{1} & 0.03 & \multicolumn{1}{c|}{41.33} & \multicolumn{1}{c|}{2} & 0.04 & \multicolumn{1}{c|}{78.14} & \multicolumn{1}{c|}{2} & 0.04 & \multicolumn{1}{c|}{143.91} & \multicolumn{1}{c|}{4} & 0.0 \\ \hline
Read-Heavy 5 & \multicolumn{1}{c|}{26.12} & \multicolumn{1}{c|}{1} & 0.03 & \multicolumn{1}{c|}{42.24} & \multicolumn{1}{c|}{2} & 0.03 & \multicolumn{1}{c|}{80.01} & \multicolumn{1}{c|}{2} & 0.04 & \multicolumn{1}{c|}{148.96} & \multicolumn{1}{c|}{4} & 0.0 \\ \hline
Read-Heavy 6 & \multicolumn{1}{c|}{27.55} & \multicolumn{1}{c|}{1} & 0.02 & \multicolumn{1}{c|}{44.49} & \multicolumn{1}{c|}{1} & 0.03 & \multicolumn{1}{c|}{84.55} & \multicolumn{1}{c|}{2} & 0.04 & \multicolumn{1}{c|}{157.9} & \multicolumn{1}{c|}{4} & 0.0 \\ \hline
Read-Heavy 7 & \multicolumn{1}{c|}{13.86} & \multicolumn{1}{c|}{N/A} & 0.06 & \multicolumn{1}{c|}{21.87} & \multicolumn{1}{c|}{N/A} & 0.09 & \multicolumn{1}{c|}{39.3} & \multicolumn{1}{c|}{N/A} & 0.1 & \multicolumn{1}{c|}{65.17} & \multicolumn{1}{c|}{N/A} & 0.0 \\ \hline
Read-Heavy 8 & \multicolumn{1}{c|}{28.25} & \multicolumn{1}{c|}{1} & 0.02 & \multicolumn{1}{c|}{45.62} & \multicolumn{1}{c|}{1} & 0.03 & \multicolumn{1}{c|}{86.77} & \multicolumn{1}{c|}{2} & 0.03 & \multicolumn{1}{c|}{162.65} & \multicolumn{1}{c|}{4} & 0.0 \\ \hline
Read-Heavy 9 & \multicolumn{1}{c|}{30.0} & \multicolumn{1}{c|}{1} & 0.02 & \multicolumn{1}{c|}{48.62} & \multicolumn{1}{c|}{1} & 0.03 & \multicolumn{1}{c|}{92.77} & \multicolumn{1}{c|}{2} & 0.03 & \multicolumn{1}{c|}{174.48} & \multicolumn{1}{c|}{3} & 0.0 \\ \hline
Read-Heavy 10 & \multicolumn{1}{c|}{30.16} & \multicolumn{1}{c|}{1} & 0.02 & \multicolumn{1}{c|}{48.91} & \multicolumn{1}{c|}{1} & 0.03 & \multicolumn{1}{c|}{93.36} & \multicolumn{1}{c|}{2} & 0.03 & \multicolumn{1}{c|}{175.64} & \multicolumn{1}{c|}{3} & 0.0 \\ \hline
Read-Heavy 11 & \multicolumn{1}{c|}{19.26} & \multicolumn{1}{c|}{5} & 0.04 & \multicolumn{1}{c|}{30.62} & \multicolumn{1}{c|}{8} & 0.05 & \multicolumn{1}{c|}{57.13} & \multicolumn{1}{c|}{15} & 0.06 & \multicolumn{1}{c|}{101.91} & \multicolumn{1}{c|}{26} & 0.0 \\ \hline
Read-Heavy 12 & \multicolumn{1}{c|}{28.97} & \multicolumn{1}{c|}{1} & 0.02 & \multicolumn{1}{c|}{46.91} & \multicolumn{1}{c|}{1} & 0.03 & \multicolumn{1}{c|}{89.32} & \multicolumn{1}{c|}{2} & 0.03 & \multicolumn{1}{c|}{167.45} & \multicolumn{1}{c|}{3} & 0.0 \\ \hline
Read-Heavy 13 & \multicolumn{1}{c|}{29.81} & \multicolumn{1}{c|}{1} & 0.02 & \multicolumn{1}{c|}{48.28} & \multicolumn{1}{c|}{1} & 0.03 & \multicolumn{1}{c|}{92.13} & \multicolumn{1}{c|}{2} & 0.03 & \multicolumn{1}{c|}{173.19} & \multicolumn{1}{c|}{3} & 0.0 \\ \hline
Read-Heavy 14 & \multicolumn{1}{c|}{30.42} & \multicolumn{1}{c|}{1} & 0.02 & \multicolumn{1}{c|}{49.44} & \multicolumn{1}{c|}{1} & 0.03 & \multicolumn{1}{c|}{94.21} & \multicolumn{1}{c|}{2} & 0.03 & \multicolumn{1}{c|}{177.39} & \multicolumn{1}{c|}{3} & 0.0 \\ \hline
Read-Heavy 15 & \multicolumn{1}{c|}{22.77} & \multicolumn{1}{c|}{2} & 0.03 & \multicolumn{1}{c|}{36.53} & \multicolumn{1}{c|}{2} & 0.04 & \multicolumn{1}{c|}{68.43} & \multicolumn{1}{c|}{4} & 0.04 & \multicolumn{1}{c|}{126.6} & \multicolumn{1}{c|}{7} & 0.0 \\ \hline
Read-Heavy 16 & \multicolumn{1}{c|}{26.54} & \multicolumn{1}{c|}{1} & 0.02 & \multicolumn{1}{c|}{42.94} & \multicolumn{1}{c|}{1} & 0.03 & \multicolumn{1}{c|}{81.52} & \multicolumn{1}{c|}{2} & 0.04 & \multicolumn{1}{c|}{152.01} & \multicolumn{1}{c|}{4} & 0.0 \\ \hline
Read-Heavy 17 & \multicolumn{1}{c|}{8.65} & \multicolumn{1}{c|}{26} & 0.05 & \multicolumn{1}{c|}{13.47} & \multicolumn{1}{c|}{40} & 0.06 & \multicolumn{1}{c|}{24.84} & \multicolumn{1}{c|}{74} & 0.08 & \multicolumn{1}{c|}{42.85} & \multicolumn{1}{c|}{127} & 0.0 \\ \hline
Read-Heavy 18 & \multicolumn{1}{c|}{29.78} & \multicolumn{1}{c|}{1} & 0.02 & \multicolumn{1}{c|}{48.31} & \multicolumn{1}{c|}{1} & 0.03 & \multicolumn{1}{c|}{91.96} & \multicolumn{1}{c|}{2} & 0.03 & \multicolumn{1}{c|}{173.15} & \multicolumn{1}{c|}{3} & 0.0 \\ \hline
Read-Heavy 19 & \multicolumn{1}{c|}{30.06} & \multicolumn{1}{c|}{1} & 0.02 & \multicolumn{1}{c|}{48.85} & \multicolumn{1}{c|}{1} & 0.03 & \multicolumn{1}{c|}{93.03} & \multicolumn{1}{c|}{2} & 0.03 & \multicolumn{1}{c|}{174.9} & \multicolumn{1}{c|}{3} & 0.0 \\ \hline
Read-Heavy 20 & \multicolumn{1}{c|}{30.31} & \multicolumn{1}{c|}{1} & 0.02 & \multicolumn{1}{c|}{49.17} & \multicolumn{1}{c|}{1} & 0.03 & \multicolumn{1}{c|}{93.88} & \multicolumn{1}{c|}{2} & 0.03 & \multicolumn{1}{c|}{176.83} & \multicolumn{1}{c|}{3} & 0.0 \\ \hline
Read-Heavy 21 & \multicolumn{1}{c|}{28.35} & \multicolumn{1}{c|}{1} & 0.02 & \multicolumn{1}{c|}{46.01} & \multicolumn{1}{c|}{1} & 0.03 & \multicolumn{1}{c|}{87.54} & \multicolumn{1}{c|}{2} & 0.03 & \multicolumn{1}{c|}{164.02} & \multicolumn{1}{c|}{3} & 0.0 \\ \hline
Read-Heavy 22 & \multicolumn{1}{c|}{20.58} & \multicolumn{1}{c|}{1} & 0.03 & \multicolumn{1}{c|}{33.3} & \multicolumn{1}{c|}{2} & 0.04 & \multicolumn{1}{c|}{63.02} & \multicolumn{1}{c|}{3} & 0.05 & \multicolumn{1}{c|}{114.34} & \multicolumn{1}{c|}{5} & 0.0 \\ \hline
Read-Heavy 23 & \multicolumn{1}{c|}{29.89} & \multicolumn{1}{c|}{1} & 0.02 & \multicolumn{1}{c|}{48.48} & \multicolumn{1}{c|}{1} & 0.03 & \multicolumn{1}{c|}{92.51} & \multicolumn{1}{c|}{2} & 0.03 & \multicolumn{1}{c|}{173.99} & \multicolumn{1}{c|}{3} & 0.0 \\ \hline
\end{tabular}}
\caption{Profiling costs, iters to earn back profiling costs,
and est. error for 15, 25, 50, and 100\% samples (G$\rightarrow$A1 1TB).  }
\label{tbl:profiling_sample}
\end{table}

\mypar{Other Complementary Optimizations} Other prior work saves money through
semantic caching and distributed query optimization techniques
~\cite{dar_semantic,crystal,dynamat,history-aware},
optimizing data placement~\cite{cache_investment,
polychroniou_distributed_2018}, and view selection and materialization in data
warehouses~\cite{nadeau_achieving_2002,
chirkova_formal_2002, automv, spectrum_perf}.
These efforts save costs within a single pricing model and can be applied to
databases \emph{prior} to our analysis across multiple pricing
models; for that reason they complement our research.

\mypar{Cloud-Agnostic Query Execution} Recent position papers have emphasized the
need to build cloud-agnostic data infrastructure.
Berkeley's Sky Computing vision outlines opportunities for multi-cloud
workload execution \cite{chasins_sky_2022}. Our work on \sys{} emphasizes cost
savings across pricing models.

\mypar{Federated Query Execution} Some prior work
    improves performance for federated queries 
    by using metadata from federated
    sources~\cite{semagrow}, by improving query planning for federated
    queries~\cite{prog_fed_q}, or by building full federated query
    systems~\cite{garlic}. These works do not aim to save money
    and consider a single execution backend and multiple
storage endpoints, while \sys{} uses multiple execution backends
with different pricing models to save money.

\section{Conclusion}
\label{sec:conclusion}

This paper presents, exploits, and evaluates two money saving opportunities for
cloud analytical workloads. The key is to schedule queries based on the
resources they consume onto \emph{beneficial} pricing models
offered by cloud vendors. 
We measure hard-to-estimate query information, implement the
inter- and intra-query algorithms, and use IaaS to save money,
all while honoring runtime constraints.

We hope this work will encourage further investigation into multi-cloud savings
opportunities. Ideally, this line of work fosters competition
between cloud vendors, driving down prices and benefiting users. Cloud vendors may, however, simply
modify prices to prevent data movement and lock-in users.
Even in extreme situations multi-cloud opportunities exist, and we
hope that cloud vendors choose to reduce costs for users and pay
for the revenue loss by becoming more energy-efficient to lower 
internal costs.

\balance

\bibliographystyle{ACM-Reference-Format}
\bibliography{bib/motivation,bib/dbs,bib/academia_cost_save,bib/industry_cost_save,bib/cache,bib/pricing,bib/recurrent_workloads,bib/est,bib/olap_dbs,bib/revision,bib/etl,bib/periodic,bib/misc}

\end{document}